\documentclass[12pt,preprint]{aastex}
\usepackage{lscape, url}
\usepackage[ps2pdf,bookmarks=true,bookmarksnumbered=true]{hyperref}

\shorttitle{IRC +10216 long term evolution}
\shortauthors{Males et al.}

\begin{document}

\title{Four decades of IRC +10216: evolution of a carbon rich dust shell resolved at $10\mu\mbox{m}$ with MMT adaptive optics and MIRAC4
\footnote{The observations reported here were partially obtained at the MMT Observatory, a facility operated jointly by the Smithsonian Institution and the University of Arizona.}
\footnote{Based in part on data from The United Kingdom Infrared Telescope, which is operated by the Joint Astronomy Centre on behalf of the Science and Technology Facilities Council of the U.K.}
\footnote{Based in part on observations with ISO, an ESA project with instruments funded by ESA Member States (especially the PI countries: France, Germany, the Netherlands and the United Kingdom) and with the participation of ISAS and NASA.}
}
\author{Jared R. Males \footnote{email: jrmales@email.arizona.edu}, Laird M. Close, Andrew J. Skemer, Philip M. Hinz, William F. Hoffmann}
\affil{Steward Observatory, Department of Astronomy, University of Arizona, Tucson, AZ 85721}

\author{Massimo Marengo}
\affil{Department of Physics and Astronomy, Iowa State University, Ames, IA  50011}

\begin{abstract}
The evolved carbon-rich AGB star IRC +10216 (CW Leo) is the brightest mid-infrared source outside the solar system, as well as one of the closest examples of an evolved star losing mass.  It has a complex and variable circumstellar structure on small scales in the near-IR, and mid-IR interferometry has revealed a dynamic dust formation zone.  We have obtained diffraction limited imaging and grism spectroscopy of IRC +10216 at the 6.5m MMT in the N-band ($\sim8-13\mu\mbox{m}$).  These new observations show that a change has occurred in the dust shell surrounding IRC +10216 over the last two decades, which is illustrated by a change in the apparent shape of the well known SiC spectral feature at $\sim11\mu\mbox{m}$ and a reduction in the continuum at $13\mu\mbox{m}$.  As expected, our diffraction limited spatial information shows an extended circumstellar envelope.  We also demonstrate that the dusty envelope appears to be $\sim30\%$ larger at the wavelengths of the SiC feature, likely due to the increased opacity of SiC.  The deconvolved FWHM of the object increases from $0.43"$ $(\sim 56$ AU) for $\lambda < 10 \mu\mbox{m}$ to $0.58"$ ($\sim 75$ AU) at $11.8\mu\mbox{m}$, then decreases to $0.5"$ ($\sim65$ AU) at $12.7\mu\mbox{m}.$  Our estimates of IRC +10216's size allow us to plausibly tie the change in the spectrum over the last 12.5 years to the evolution of the dusty circumstellar envelope at speeds of 12-17 km sec${^{-1}}$.
\end{abstract}

\section{Introduction}
\subsection{The carbon star IRC +10216}
When stars of low to intermediate mass are in the last stages of nuclear burning on the asymptotic giant branch (AGB) of the Hertzsprung-Russel (HR) diagram, they are typically characterized by high luminosity, which varies with long periods ($1-2$yrs), and mass loss.  The high mass loss rates, up to $\dot{M} \sim 10^{-4} M_{\sun} \mbox{ yr}^{-1}$, ultimately lead to the termination of nuclear burning and produce dusty, often optically thick, circumstellar envelopes (CSEs) which provide one of the key observational features of AGB stars.  Of particular concern here, the CSE is often the dominant source of light in the near and mid-infrared (IR) and contains much information about the evolution and mass loss history of the enshrouded star.  For thorough treatments of AGB stars and their evolution see \citet{habingAGB} and \citet{2005ARA&A..43..435H}, and references therein.

The carbon star IRC +10216 (CW Leo) is perhaps the best studied example of an AGB star.  Since its discovery by the 2.2 $\mu\mbox{m}$ survey in 1969  \citep{1969ApJ...158L.133B}, IRC +10216 has been recognized as a star enshrouded by a thick CSE.  It exhibits large ($>2$X) changes in luminosity over its 649 day cycle \citep{1992A&AS...94..377L} and is extremely bright in the mid-IR ($>10^4$ Jy).  IRC +10216 is classified as a carbon star \citep{1970ApJ...162L..15H}, implying that the ratio of carbon to oxygen in its photosphere is greater than 1.  It is believed to be in the final transitional stage between the thermal pulse (TP) AGB and the post-AGB/planetary nebula stage \citep{1998MNRAS.300L..29S,2000A&A...357..169O}.  More recently \citet{2001Natur.412..160M} reported the detection of warm water vapor in the CSE of IRC +10216, and \citet{2010Natur.467...64D} have reported the detection of many water lines in the CSE by the Herschel satellite \citep{2010A&A...518L...1P}.

IRC +10216 is clearly a fascinating and well studied object, and it is impossible to fully review the extensive literature on it here.  As such we focus mainly on the N band atmospheric window, which is bounded by water vapor at $\lambda \lesssim 8\mu\mbox{m}$ and $CO_2$ at $\lambda \gtrsim 14\mu\mbox{m}$, and on high-spatial resolution imaging and interferometry of the CSE.  In carbon stars, the N-band spectrum usually shows the emission feature of SiC.

\subsection{SiC dust}
\label{sect:intro_sic}
Around 40 years ago the production of SiC \citep{1969Phy....41..139F, 1969ApJ...155L.185G} and the presence of its emission feature near $11\mu\mbox{m}$ \citep{1971Natur.229..237G} were predicted.  This feature was then discovered in the N band spectra of carbon stars \citep{1972A&A....21..239H}, including in IRC +10216 by \citet[hereafter TC74]{1974ApJ...188..545T}.  In this paper we present observational evidence that either the spectroscopic SiC feature in IRC +10216, or the underlying continuum, has undergone a significant change in the last 15 to 20 years, so we will briefly discuss some of the previous work attempting to connect the properties of this feature to the evolutionary state of the underlying AGB stars.  The Infrared Astronomical Satellite Low-Resolution Spectrometer (IRAS/LRS) provided a wealth of data in the mid-IR spectral region, including a catalog of AGB star spectra.  These data have been used extensively to study the $\sim11\mu\mbox{m}$ SiC feature\footnote{We adopt the nomenclature ``$\sim11\mu\mbox{m}$'' of \citet*{2005ApJ...634..426S} to indicate the varied peak wavelengths of this feature.} of carbon stars, generally finding a positive correlation between dust continuum temperature and the strength of the emission peak \citep{1987A&A...186..271B, 1990A&A...237..354C, 1998AJ....115..809S}.  

A common feature of these efforts has been an attempt to relate the SiC feature and other characteristics of the mid-IR dust spectra to the long term evolution of the host AGB star.   \citet{2006ApJ...652.1654T} provide a useful review of this work, and use the more recent Infrared Space Observatory \citep[ISO, ][]{1996AA...315L..27K} Short Wavelength Spectrometer \citep[SWS, ][]{ISO_V_SWS} data set to further investigate correlations between the SiC peak strength, peak wavelength, and dust continuum temperature.  They ultimately conclude that there are no useful correlations, and blame poor continuum fitting for the previous results.  

\subsection{IRC +10216 in the spatial domain}
\label{sect:spatial_domain}
IRC +10216 has also provided many fascinating results in the spatial domain. Given its extreme brightness in the mid-IR, it was an early target for interferometry, and has more recently been subject to intense study in the near-IR.  At wider spatial scales, visible wavelength imaging has shown an extended dusty envelope composed of multiple shells.  We will now briefly review some of these results, with particular interest in their implications for the process and variability of mass loss from IRC +10216.

Deep optical observations have shown that IRC +10216 is surrounded by multiple dusty shells, which can be seen scattering ambient galactic light out to separations of $\sim 200 "$.  \citet{1999A&A...349..203M} analyzed these shells in B and V band images from the Canada France Hawaii Telescope (CFHT) on Mauna Kea and concluded that some process modulates the mass loss on a timescale of 200-800 years, and later found evidence for timescales as short as 40 years \citep{2000A&A...359..707M} using Hubble Space Telescope (HST) imaging.  \citet{2006A&A...455..187L} used deep Very Large Telescope (VLT) V band images to show that the shells can be resolved into even smaller structures.  These shells appear to be only approximately spherical and are azimuthally incomplete, indicating that the mass loss is not isotropic.  HST imaging of the inner $\sim10"$ shows a nearly bipolar structure, reminiscent of the typical but poorly understood structure of planetary nebulae \citep{1998MNRAS.300L..29S}.

High spatial resolution observations in the near-IR have produced a fascinating picture of the inner portions of the dusty envelope around IRC +10216 (which are invisible in the optical). Using speckle-masking interferometry in the K' band \citet{1998A&A...333L..51W} found the inner $1/2"$  to be composed of at least 5 distinct clumps, indicating an inhomogeneous recent mass loss history.  \citet{1998A&A...334L...5H} then presented diffraction limited imaging data which showed that between 1989 and 1997 these clumps had undergone significant evolution, exhibiting relative motion and some either appearing or becoming brighter.  \citet{2000ApJ...543..284T} showed significant relative motion of various components of the dust, with possible acceleration, based on 7 epochs of sparse aperture mask interferometric imaging in K band using the Keck I telescope\footnote{See the movie: \url{http://www.physics.usyd.edu.au/~gekko/irc10216.html}}.  Interestingly these authors found no evidence for new dust production during these observations.

Relative motion within the inner regions of the dust shell were also found by \citet{2000A&A...357..169O}, who argued that this evolution was not related to the $\sim 2$ year luminosity cycle in any simple way.  In a related effort, extensive radiative transfer modeling was conducted by \citet{2001A&A...368..497M} taking into account much of the archival multi-wavelength data set (including spectral and spatial information).  \citet{2002A&A...392..921M}  used their model to explain the time evolution reported by \citet{2000A&A...357..169O}. A key conclusion from this study is that since its discovery IRC +10216 has been undergoing an intense period of mass loss, probably starting $\sim50$ years earlier.  They also concluded that the mass loss rate had recently increased.

IRC +10216 has been repeatedly studied by interferometers in the mid-IR.  \citet[hereafter MHL80]{1980ApJ...235L..27M} measured visibilities at 2.2, 3.5, 5.0, 8.4, 10.2, 11.1, 12.5 and $\sim20 \mu\mbox{m}$, at several epochs and position angles (PAs).  They found evidence for asymmetry at the short wavelengths, indicating an elongation along PA $\sim25^o$, which matched the early optical images and has been confirmed repeatedly by later observations \citep{1998MNRAS.300L..29S, 2006A&A...455..187L}.  No evidence of this elongation was found at 11.1$\mu\mbox{m}$ however.  It was also noted that apparent size, but not morphology, changes with photometric phase.

This object has also been observed at $\sim11\mu\mbox{m}$ using the UC Berkeley Infrared Spatial Interferometer (ISI).  \citet{1990ApJ...359L..59D} generally confirmed the large change in visibilities with photometric phase found by MHL80, and argued that dust was being formed much closer to the star than previous studies had found.  Using data from the ISI taken $\sim10$ years later, \citet{2000ApJ...543..861M} found that the inner radius of the dust had moved away from the star.  This result was based on model fits to the visibilities, and led them to conclude that no new dust was being formed for most of the 1990's.  This appears to contradict the radiative transfer based mass loss predictions of \citet{2002A&A...392..921M}.  Most recently the ISI detected some asymmetry at 11.15$\mu\mbox{m}$ using baselines of up to 12m \citep{2007ApJ...657.1042C}.

\subsection{New results from the MMT}

Here we present new spatially resolved mid-IR photometry and spectroscopy of IRC +10216 with high resolution, AO corrected, spatial information, obtained at the MMT on Mt. Hopkins, AZ, in 2009 and 2010.  We first describe our observations and data reduction, paying particular attention to the correction needed when observing an extended object with a spectroscopic slit and a diffraction limited beam.  We also review nearly four decades of measurements of the spectrum of IRC +10216.  We then discuss our new results in context with the previous work on IRC +10216.

\section{Observations and data reduction}
We observed IRC +10216 at two epochs separated by approximately 1 year, using the 4th generation Mid-Infrared Array Camera (MIRAC4), fed by the MMT Adaptive Optics (MMTAO) thermally efficient adaptive secondary mirror \citep{2003SPIE.5169...17W}.  Infrared light first passes through the Bracewell Infrared Nulling Cryostat (BLINC, \citet{2000SPIE.4006..349H})\footnote{Further information on MIRAC4 and BLINC can be found at \url{http://zero.as.arizona.edu/miracblinc}}, with visible light being reflected to the visible wavelength wavefront sensor of the AO system.  This system routinely achieves $\sim 98\%$ Strehl ratios at $10\mu\mbox{m}$ \citep{2003ApJ...598L..35C}, and can super-resolve structure smaller than its diffraction limit \citep{2006ApJ...647..464B, 2008ApJ...676.1082S}.  In addition to imaging, MIRAC4 has a grism spectroscopy mode described in \citet{2009PASP..121..897S}.

\subsection{2009 bandpass photometry}
We observed IRC +10216 on 13 Jan 2009 UT with the imaging mode of BLINC/MIRAC4 using its fine plate scale (0.055 arcsec/pixel).  Conditions were photometric, with excellent seeing, estimated to be better than FWHM=0.5" at V from the AO acquisition camera.  To avoid saturation from the extremely bright source ($\sim$40,000Jy) we read out MIRAC4's array with a 0.008s frame time.  IRC +10216 is optically faint (R mag $>$ 15), and we were unable to close the MMTAO loop.  As a result these observations were taken with the adaptive secondary in its static position, which uses a pre-determined set of actuator commands to hold the mirror shape.  We took data in the typical fashion for MIRAC4: chopping using the BLINC internal chopper and telescope nods in the perpendicular direction.  In the case of IRC +10216 we set the nod amplitude to be large enough that only one pair of chops was on the detector since the object was expected to be significantly extended.  Observations of the standard star $\mu$ UMa were taken immediately after IRC +10216 in identical fashion,  but for $\mu$ UMa the nod amplitude was set so that all four positions were on the detector to increase observing efficiency. Table \ref{tab:obslog} lists the filters and airmasses for these observations.

The data were reduced by first applying a custom artifact removal script developed for the MIRAC4 detector \citep{2008ApJ...676.1082S}, which also performs the background subtraction of the chop-nod sets.  Each frame was then inspected to look for bad chops (caused by the chopper sticking) and excessive pattern noise from the detector.  Frames with these problems were discarded.  Photometry was conducted on the individual images, rather than registering and combining, to allow an empirical estimation of the uncertainties from the artifact reduction and background subtraction processes.  We used the DAOPHOT package in IRAF, and selected the best photometric aperture for the standard and object based on the mean ``curve of growth'' for counts vs. aperture radius.  Since IRC +10216 is extended, the aperture where the curve flattened was always much wider than for the PSF.  

Following \citet{2010ApJ...711.1280S} we applied a telluric correction to the photometry using transmission curves provided by Gemini Observatory\footnote{\url{http://www.gemini.edu/sciops/telescopes-and-sites/observing-condition-constraints/}} calculated with the ATRAN model atmosphere code \citep{1992NASATM},  and an estimate of 3mm precipitable water vapor (PWV) and the airmass of the observations.  The PWV assumption is supported by contemporaneous PWV measurements taken on Kitt Peak ($74$ km west-northwest), and as noted in \citet{2010ApJ...711.1280S} the correction at these wavelengths is generally insensitive to PWV.  In the $9.79\mu\mbox{m}$ filter the correction was $+2.5\%$ (due to telluric ozone), and in all others it was $<1\%$.  Finally, we normalized the photometry by the \citet{1996AJ....112.2274C} flux for $\mu$ UMa.  The results are presented in Table \ref{tab:phot}.  

The uncertainty in our photometry was calculated in similar fashion to that used for grism spectroscopy in \citet{2010ApJ...711.1280S} and below (Section \ref{sect:grism}), with the exception that for bandpass photometry we did not assume a correlated global uncertainty.  As discussed above we performed photometry on individual frames, which provides an empirical measurement uncertainty for both the object and standard.  Thus our measurement uncertainty includes the random effects of detector artifacts and our removal procedure.  We then add in quadrature the mean uncertainty in the $\mu$ UMa standard flux across the filter bandpass from \citet{1996AJ....112.2274C}, and use the telluric calibration uncertainties from \citet{2010ApJ...711.1280S}, which were measured the following night.  The values used and the final total $1\sigma$ uncertainty are included in Table \ref{tab:phot}.

When compared to our (normalized) grism spectrum from a year later, the photometry from 9.8$\mu\mbox{m}$ to 12.5$\mu\mbox{m}$ matches very well.  In the 8.7$\mu\mbox{m}$ filter, however, there is a $\sim 30\%$ discrepancy between the bandpass photometry and the grism data, as well as with archival data.  We are suspicious of this data point since it represents a high counts regime of the detector not well understood, but we do not yet have any specific reason to discard it.  We discuss this further in Appendix A.

\subsection{2010 grism spectroscopy}
\label{sect:grism}
We observed IRC +10216 nearly one year later on 1 Jan 2010, UT in the grism spectroscopy mode of MIRAC4, using a 1" slit.  With this configuration MIRAC4 has a spectral resolution of $R\sim125$ and a spatial resolution of $\lambda/D\sim0.32"$ at $10\mu\mbox{m}$.   The detector wavelength scale was calibrated at the telescope using a well characterized polystyrene sample and fitting a quadratic function to the measured centroids of features in the spectrum.  The coarse platescale used for grism work was measured using the binary $\alpha$ Gem on 2 Jan 2010, UT and elements from the USNO Sixth Catalog of Orbits of Visual Binary Stars\footnote{\url{http://ad.usno.navy.mil/wds/orb6.html}} \citep{2001AJ....122.3472H}.  We found a value of $0.107$"/pixel.   

Conditions on 1 Jan 2010 UT were photometric.  Through a combination of excellent seeing conditions and IRC +10216 being near its brightness maximum, we were able to lock the MMTAO system on IRC +10216 with a loop speed of 25Hz.  We set the frame time to 0.008s, and to ensure that we could take advantage of the diffraction limited information being delivered by the MMTAO system we read out each 0.008s frame.  Due to its extreme brightness at $10\mu\mbox{m}$ only a few of these short frames were needed to provide sufficient S/N, and this data taking mode allows us to reject frames with bad slit alignment due to residual tip/tilt errors and frames with excessive artifacts.

Observations of the standard $\mu$ UMa were challenging for nearly opposite reasons.  Ordinarily one tries to operate the AO system with identical parameters between PSF and science object, but in the optical $\mu$ UMa saturated the wavefront sensor (WFS) at speeds slower than 100Hz.  At $10\mu\mbox{m}$ $\mu$ UMa is a factor of $\sim500$ fainter (even though it is one of the brightest 10$\mu\mbox{m}$ standards), so longer integrations are required to efficiently build S/N.  Table \ref{tab:obslog} lists the details of these observations.

Determining Strehl ratio for these observations is problematic.  We did not take data in the imaging mode (other than for slit alignment) because of the limited time ($\sim1$ hr) that conditions were good enough to lock the MMTAO system on this faint star.  Without two dimensional imaging data it is difficult to directly measure Strehl ratio from our PSF observations.  In addition, since we necessarily operated the AO system with different parameters, any such measurement would not apply to IRC +10216.  At 10$\mu\mbox{m}$ the dominant wavefront error term will be from loop delay (servo error), even on an optically faint target such as IRC +10216.  Based on the very high Strehl ratios routinely achieved by MMTAO and MIRAC4 (98\%), the WFS integration times (40ms), and our use of short exposures, we estimate the Strehl ratio of our IRC +10216 observations to be $\sim80\%$. 

The Moon moved closer to IRC +10216 on the following night (2 Jan) and seeing was somewhat worse, so we were unable to lock MMTAO on IRC +10216.  Though we took seeing limited data, we find that the good spatial information provided by AO is necessary to adequately correct for differential slit loss between the point source standard and a resolved IRC +10216.  We did, however, take AO-on spectra of the standards $\mu$ UMa and $\beta$ Gem, which we use to calibrate our slit loss correction procedure.  Details of these observations are also included in Table \ref{tab:obslog}.

\begin{figure}
\begin{center}
\includegraphics[scale=.8]{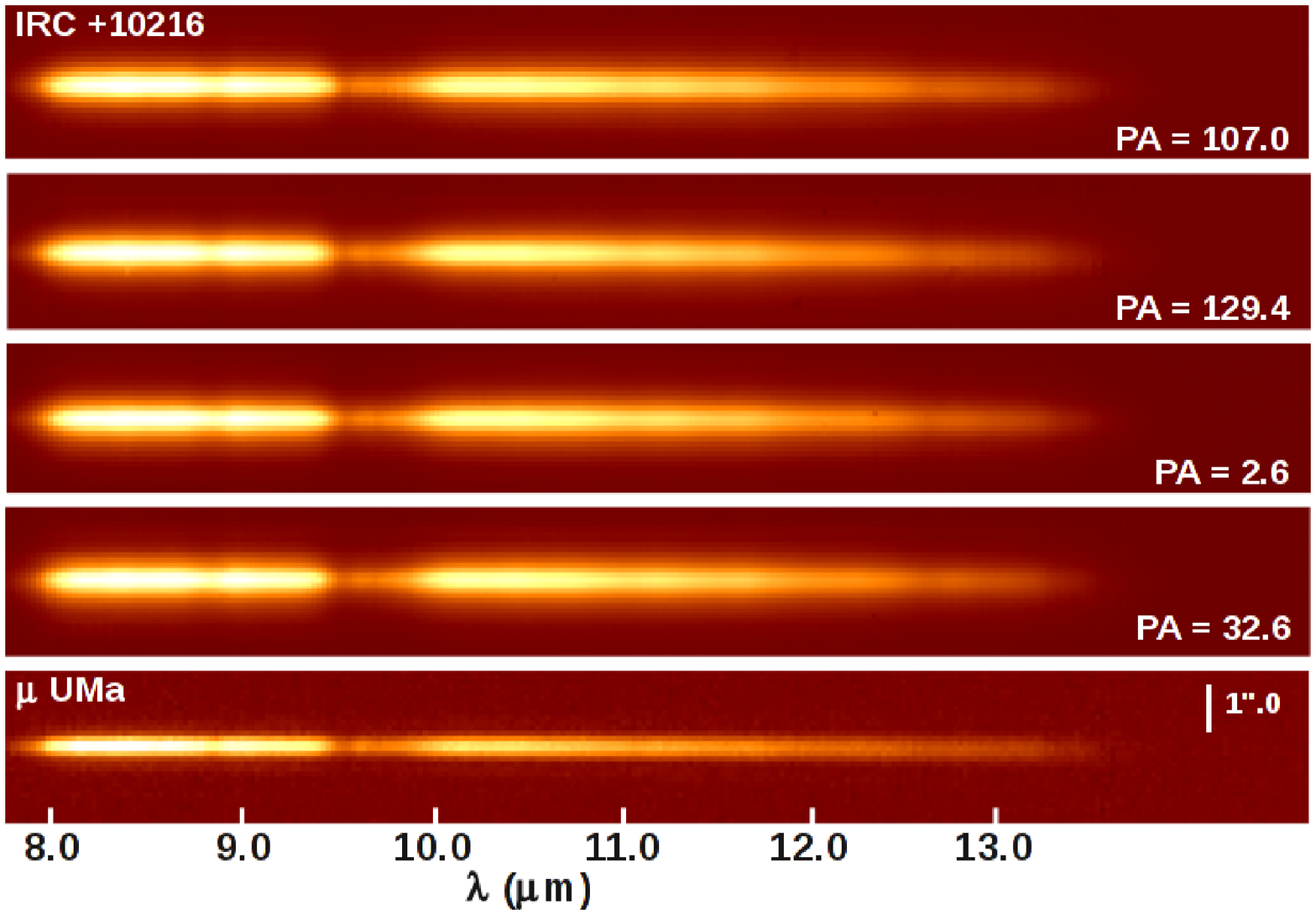}
\end{center}
\caption{MIRAC4 grism observations of IRC +10216 and $\mu$ UMa with the MMTAO loop closed.  Here we present $\sim0.3"$ diffraction limited spatial information in the vertical direction and $R\sim125$ spectral information in the horizontal.  See Table \ref{tab:obslog} and the text for the details of the observations, especially AO system parameters which were necessarily different due to the relative optical brightness of the two sources.  Compared to the PSF, IRC +10216 is clearly resolved.  See Figures \ref{fig:profiles} and \ref{fig:fwhm} for the results of extracting profiles in the spatial direction and Figure \ref{fig:grism_phot} for the fully reduced spectrum of IRC +10216.  Note the impact of telluric ozone absorption between $9$ and $10\mu\mbox{m}$, and the decreasing sensitivity starting at 13$\mu\mbox{m}$.
\label{fig:grism_obs}}
\end{figure}

Reduction of grism data is similar to the imaging procedure described above, except that our data were taken with nods only, as the chopper is unnecessary for very bright sources and can sometimes cause slit-misalignment.   We used the same artifact removal script, and the images at each position angle were registered and median combined.  Our fully reduced images of IRC+10216 and $\mu$ UMa are presented in Figure \ref{fig:grism_obs}.

\begin{figure}
\begin{center}
\includegraphics[angle=90,scale=.68]{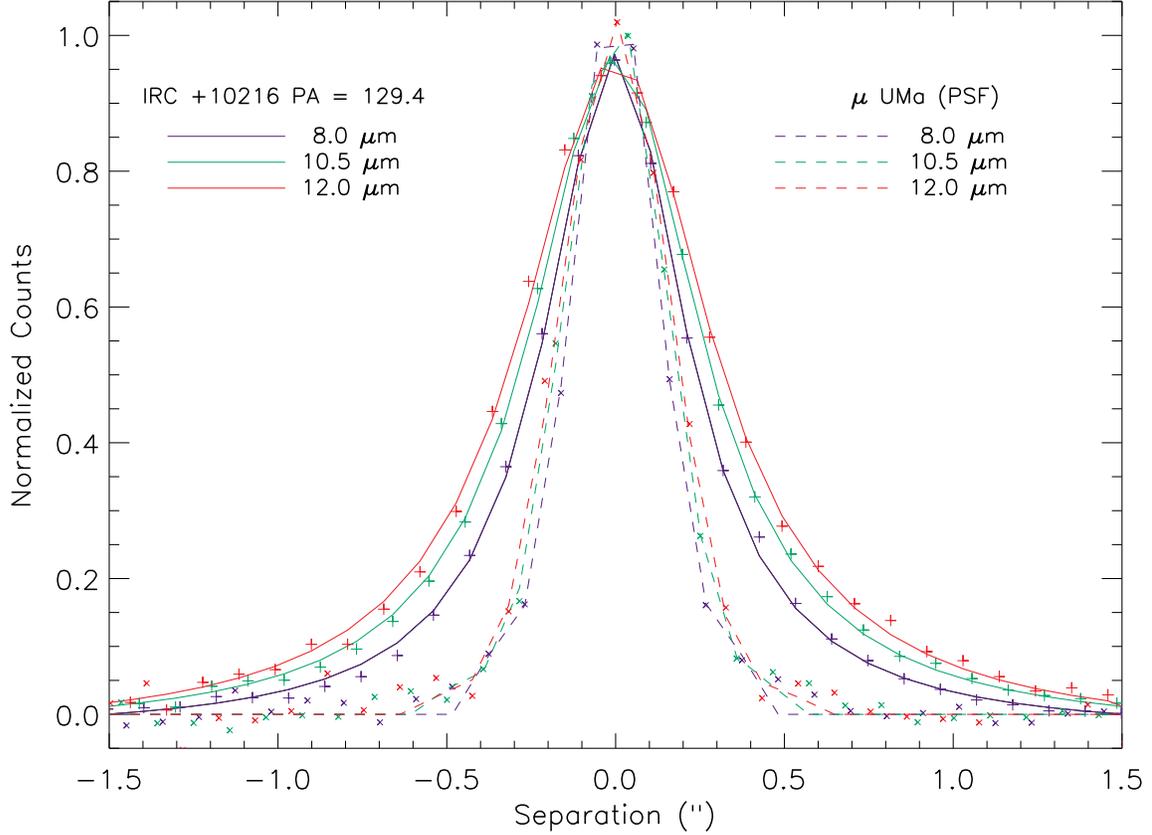}
\end{center}
\caption{Normalized spatial profiles of IRC +10216 and the PSF standard $\mu$ UMa, at three discrete wavelengths (i.e. single detector rows) for a single position angle (129.4).  The data are denoted by x's for $\mu$ UMa and +'s for IRC +10216.  The PSF core is well fit by a Gaussian (dashed lines), as expected for a well corrected Airy disk (we don't fit past the first airy minimum, which can be seen along with the first airy ring at $\sim 0.5$").  IRC +10216 is well described by a Lorentzian profile (solid lines), though there are apparent correlated discrepancies at wider separations.  This result, and similar results for the other position angles, gives us confidence that IRC +10216 is resolved and the FWHM determined by fitting a Lorentzian gives a meaningful proxy for object size vs. wavelength.
\label{fig:profiles}}
\end{figure}

\begin{figure}
\begin{center}
\includegraphics[angle=90,scale=.68]{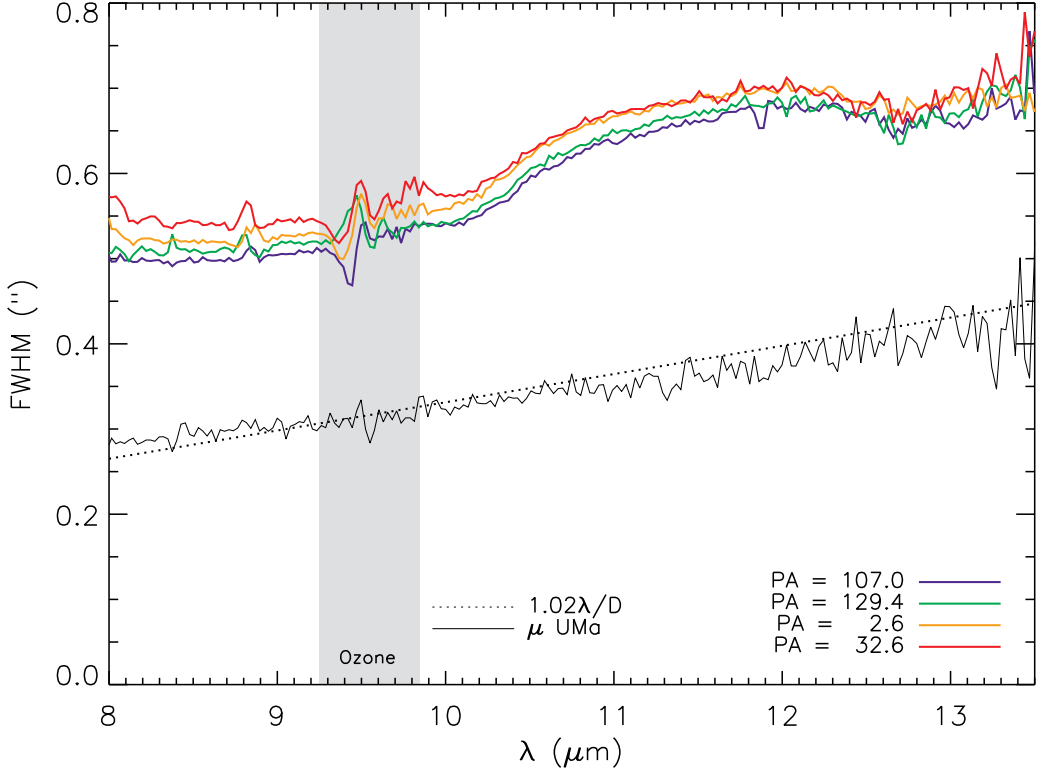}
\end{center}
\caption{The results of fitting profiles of the images presented in Figure \ref{fig:grism_obs} as a function of wavelength, plotted as FWHM.  The PSF core was fit with a Gaussian, which is expected to match a well corrected Airy pattern inside the first Airy minimum.  For comparison we plot the predicted result for a circular pupil 6.35m in diameter with an 11\% central obscuration (i.e. the MMT with the adaptive secondary, dotted line).  Though the slope of the line does not match perfectly (likely due to a stop reducing the effective diameter or changing the central obscuration) it shows that the MMTAO system reached the diffraction limit for these observations.  IRC +10216 was fit with a Lorentzian, which, though chosen for no astrophysical reason, matches our data well.  The fits show clear evidence of a size change with wavelength, distinct from the effect of diffraction, between 10.2 and 12.6 $\mu\mbox{m}$, which matches the SiC emission feature.  To avoid confusion we have indicated the spectral region typically impacted by telluric ozone.  Also note the small feature at $\sim8.8\mu\mbox{m}$ which can be attributed to a sharp feature in the detector QE.
\label{fig:fwhm}}
\end{figure}

Our first step in analyzing the data was to fit the spatial profiles of the  PSF and IRC +10216 at each detector row.  We found that a Lorentzian is a good fit for IRC +10216 out to wide separations from the peak, generally achieving $\chi^2_{\nu} < 2$ across the entire wavelength range.  As expected a Gaussian was good for the core of the Airy pattern of our PSF standard.  We used profile plots (an example of which is shown in Figure \ref{fig:profiles}) to assess the quality of this analysis.  The chosen functions describe the core of the objects well, and we find that the Lorentzian full-width at half-maximum (FWHM) is a meaningful proxy for the size of IRC +10216 relative to the PSF.  Based on this conclusion we show FWHM vs. wavelength in Figure \ref{fig:fwhm}, where we see that the PSF was essentially diffraction limited.  The comparison is not perfect due to the difference in AO system parameters between the two objects, but it is clear that IRC +10216 is extended.  It is also apparent that the dependence of size on wavelength is much more complicated than mere $\lambda/D$ scaling due to diffraction.  

\begin{figure}
\begin{center}
\includegraphics[angle=90,scale=.68]{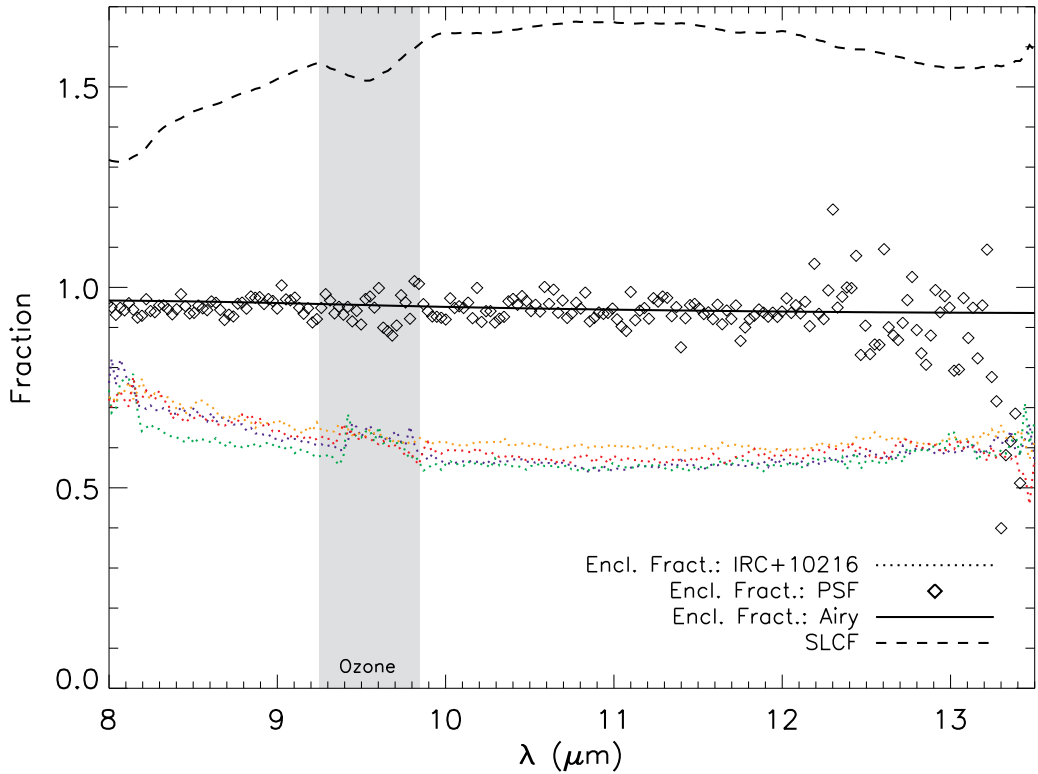}
\end{center}
\caption{Slit-loss correction calculations.  We show the empirically calculated flux enclosed by the slit and photometric aperture for IRC +10216 (colored dots, using the same colors as Figure \ref{fig:fwhm} to denote PA), and the median of the 5 AO-on PSF standards obtained (diamonds), four of which are from the night after the IRC +10216 data were taken.  Also plotted are the expected results for a centrally obscured Airy pattern, which we use for our final correction factor calculation to avoid introducing noise in our spectrum.  Finally, we show the resultant slit-loss correction factor (SLCF), which we multiply with the spectrum of IRC +10216.
\label{fig:sloss}}
\end{figure}

Regarding data reduction, an important conclusion to draw from Figure \ref{fig:fwhm} is that one cannot simply divide by a point source standard to calibrate this extended object when using a slit, as a different amount of light is lost due to the slit, and this effect depends on wavelength in a non-analytic way.  To quantify the effect of the slit we constructed a surface of revolution for IRC +10216 at each wavelength using the 1-D spatial profile.  We then calculated the fraction of flux enclosed by the $1"$ slit, taking into account the width of the aperture used to extract flux at each wavelength.  The results of these calculations are shown in Figure \ref{fig:sloss}, for each position angle.

For the PSF, in addition to the photometric standard taken on 1 Jan, we used the AO-on standard observations from the following night in order to improve S/N.  Each standard was analyzed independently, then we took the median of the results at each detector row (i.e. wavelength).  We compare the outcome of this procedure to that expected based on the theoretical Airy pattern for the MMT, which we processed in similar fashion, in Figure \ref{fig:sloss}.  

To correct for the differential slit-loss, we calculate the slit-loss correction factor (SLCF) as the ratio of the enclosed flux of the PSF to that of IRC +10216.  We use the theoretically calculated curve for the actual aperture due to the relatively noisier empirical results for the PSF.  We use the average of the IRC +10216 results to suppress noise, ignoring the small possible source asymmetry highlighted by our FWHM curves since it will cause only a small difference in the results.  The resultant SLCF curve is shown in Figure \ref{fig:sloss}. 

\begin{figure}
\begin{center}
\includegraphics[angle=90,scale=.68]{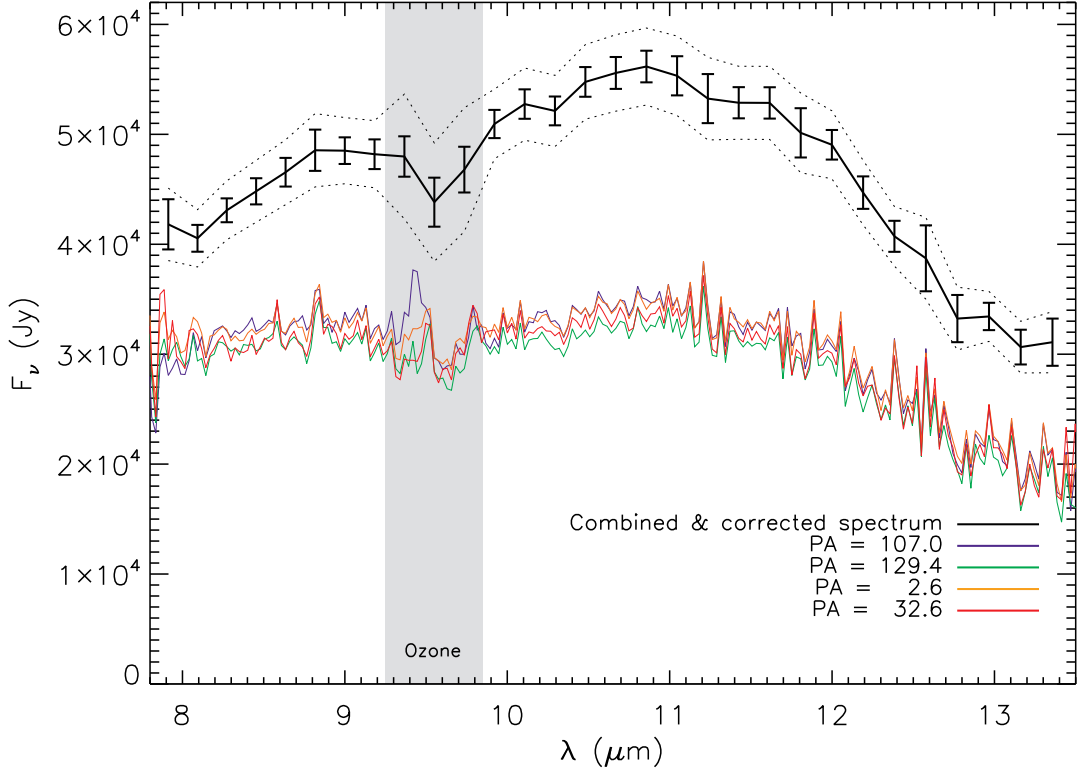}
\end{center}
\caption{Calibrated flux before and after correction for differential slit loss.   The lower curves show the raw calibrated flux, before applying the SLCF, for each of the four slit position angles.  The top curve is our fully corrected median combined spectrum, which takes into account the differential slit loss of the extended object compared to the PSF standard.  See Figure \ref{fig:sloss} and the text for further discussion of the SLCF.  The error bars denote the local error, and the dashed lines denote our total uncertainty, which in addition to the local error includes the global (correlated) uncertainties.  The average flux from $8-13\mu\mbox{m}$ is 47611 Jy, which is very similar to the value of 47627 Jy obtained by \citet{1998ApJ...502..833M} at similar phase (near maximum brightness).
\label{fig:grism_phot}}
\end{figure}

We performed aperture photometry on IRC +10216 and the standard.  As with the bandpass photometry we corrected for airmass and used the \cite{1996AJ....112.2274C} standard spectrum of $\mu$ UMa to calibrate the results.  In the PA=107.0 spectrum we found a $-1.7\% / \mu\mbox{m}$ slope compared to the other three, likely due to a slight offset in the slit or possibly a period of worse AO correction.  This slope was removed.  We show the raw spectrum at each position angle prior to applying the SLCF in Figure \ref{fig:grism_phot}.  Most of the noise in the spectrum appears to be correlated noise, i.e. it is identical in all four position angles, implying it comes from the standard.  A simple Poisson noise calculation indicates that we should have achieved slightly greater S/N in the $\mu$ UMa observation.  The higher noise in the standard is most likely due to its lower flux relative to the background and the interaction of this with the MIRAC4 detector artifacts.

The SLCF was applied to each PA, and then we rebinned by 7 pixels using the median as in \cite{2010ApJ...711.1280S}.  Though this sacrifices some spectral resolution, it has the benefit of increasing the S/N in each bin and allowing a robust empirical estimate of the uncertainties.  We used the same prescription for calculating local measurement error as \cite{2010ApJ...711.1280S}, estimating the Gaussian $1\sigma$ error from the 2nd and 6th ordered values in each bin.  The global bias (correlated error) reported by \cite{1996AJ....112.2274C} for $\mu$ UMa is negligible, so we add their total uncertainty in quadrature to the measurement error to calculate the total local uncertainty.

Global telluric error was estimated from the four spectra, which were taken at different airmasses.  We find results similar to \cite{2010ApJ...711.1280S}: $2.7\%$ outside the ozone feature and $10\%$ inside.  We do not include a separate local telluric error as this will be included in our measurement error.  Finally we adopt a $5\%$ global systematic error term from the SLCF procedure which is based on the scatter in the IRC +10216 enclosed fraction results.

The fully calibrated and slit-loss corrected results are listed in Table \ref{tab:grismspect} and shown in Figure \ref{fig:grism_phot}, along with the local and total uncertainties.  The effect of the SLCF can be seen, in addition to the overall increase in flux the SLCF has effects on the shape of the spectrum compared to the uncorrected curves.  It highlights an apparent ``bump'' at $\sim9\mu\mbox{m}$.  The SLCF also reveals a steeper negative slope longer than $\sim11\mu\mbox{m}$.  This slope matches the bandpass photometry from 2009 very well, giving us confidence in our slit loss correction procedure.  We discuss these features in more detail in section \ref{sect:discussion}.

\section{Archival data}
\label{sect:archive}
Since its discovery, IRC +10216 has been observed many times, at nearly every wavelength available to astronomers.  In this paper we concentrate mainly on the region of N-band accessible through the Earth's atmosphere.  Some of the earliest observations at these wavelengths were of IRC +10216, and it has been observed regularly over the last four decades, though, ironically, increases in telescope size and detector sensitivity may be curtailing this somewhat due to the dynamic range required to avoid saturation.   It would be impossible to account for all of the work done on this object.  Here we use a sample of N-band spectra, and several datasets of bandpass photometry.

We present these data in the context of the light curve parameters of \citet{1992A&AS...94..377L}, where the period P = 649 days, and phase $\phi =0$ at JD 2447483 (where $\phi$ varies from 0 to 1).  Based on the spread reported in the various filters used, and other determinations (e.g. 638 days in \citet{1991AJ....102..200D}), the period is uncertain by $\sim10$ days.  This should be kept in mind when comparing widely separated measurements, i.e. nearly 21 cycles have occurred between the TC74 data and our 2010 measurement so the relative phase between them could be off by 30\% or more.  This is less of a concern for more closely spaced data and we are not attempting a light-curve analysis here, rather we claim that the $\sim2$ year Mira variability isn't the source of the changes we discuss.  For our observations the star's luminosity was at $\phi = 0.27$ on 13 Jan., 2009 and $\phi=0.89$ on 1 Jan., 2010, assuming \citet{1992A&AS...94..377L}'s parameters.

\subsection{Introduction to the spectral datasets}
Here we collect a sample of N-band spectra, choosing some of the earliest measurements, two space-based observations (IRAS/LRS and ISO/SWS), and a set of observations taken on the same instrument (CGS3 at UKIRT) repeatedly over a short period of time.  We briefly describe these datasets here and any processing we did.  It is worth noting that none of these observations are affected by slit loss, as they either used no slit or had large apertures.

\subsubsection{The spectrum of Treffers and Cohen}

The spectrum of TC74 was taken with a scanning Michelson interferometer (i.e. a fourier transform spectrometer (FTS)) on a 2.2m telescope on Mauna Kea on 15 and 16 Feb., 1973 ($\phi = 0.13$).   They used the Moon as a telluric standard, and reported in arbitrary flux per unit wavenumber ($F_{\nu}$) with a resolution of 2cm$^{-1}$ ($\sim 0.02\mu\mbox{m}$).  The gap in the spectral fragments was in the original data, and though not commented on by TC74 is almost certainly due to the telluric ozone feature.  Below $\sim 8\mu\mbox{m}$ the spectrum appears to be unreliable due to telluric water vapor.

\subsubsection{The Spectrophotometry of Merrill and Stein}

\citet[hereafter MS76]{1976PASP...88..294M} used a circular variable filter (CVF) photometer to observe IRC +10216 from Mt. Lemmon, AZ.  The data are the average of observations taken between 30 Apr. and 2 May, 1974 ($\phi=0.81$) (K. M. Merrill, 2011, private communication).  We extracted the data from their Figure 2, and converted from $\lambda F_\lambda$ units to arbitrary $F_\nu$.

\subsubsection{The IRAS LRS Spectrum} 

The IRAS LRS spectrum for IRC +10216 was extracted from a database maintained by Kevin Volk\footnote{\url{http://www.iras.ucalgary.ca/~volk/getlrs_plot.html}}.  We spliced the blue and red fragments together and applied a correction for the spectral shape of the IRAS standard $\alpha$ Tau using the procedure of \citet{1992AJ....104.2030C}.  It is not possible to assign a single epoch to IRAS observations,  so we adopt the range 1 Feb. to 1 Nov. 1983 ($\phi = 0.74 - 0.16$).

\subsubsection{The ISO/SWS Spectrum}

We retrieved the reduced ISO/SWS observation of IRC +10216 from the ISO archive\footnote{\url{http://iso.esac.esa.int/}}, taken on 2 June 1996 ($\phi = 0.24$).
The ISO data presented in this paper are  from the Highly Processed Data Product (HPDP) set called `High resolution processed and defringed SWS01s', available for public use in the ISO Data Archive \citep{ISO_SWS_technote52}.  The data we are interested in span detector bands 2C (7.0-12.0$\mu\mbox{m}$) and 3A (12.0-16.5$\mu\mbox{m}$) \citep{ISO_V_SWS}. Though the pipeline attempts some defringing in band 3, we applied a 0.01$\mu\mbox{m}$ binning (averaging) to the data to reduce fringing, which is especially prominent in band 2C.  This spectrum has been published previously by \citet{1999ApJ...526L..41C} who used an earlier reduction pipeline and did not discuss the 8 to 13 $\mu\mbox{m}$ region.

\subsubsection{Spectra from UKIRT}

\citet*{1998ApJ...502..833M} (hereafter MGD98) obtained spectra of IRC +10216 at 4 epochs from 1994 to 1996 as part of a survey of variability in late type stars, using the Cooled Grating Spectrometer 3 (CGS3) at UKIRT on Mauna Kea.  Several stars were used as photometric standards at each epoch.  We considered these measurements separately, and also averaged the 4 spectra for comparison with our data, first normalizing each to 10.55$\mu\mbox{m}$ and using 0.1$\mu\mbox{m}$ bins to compensate for the slight changes in wavelength scale between observations.  The luminosity phases of these spectra were $\phi = 0.38, 0.47, 0.96,$ and $0.27$.  MGD98 noted some small fluctuations in the spectral slope (across the 10$\mu\mbox{m}$ window) with phase and though they speculated that the changes were due to a rapidly changing dust condensation zone, they could not rule out poor calibrations as the cause.  We note that their data from 22 June 1996 matches the 2 June 1996 ISO/SWS data fairly well.

\subsection{Bandpass photometry archives}
Bandpass photometry can provide a useful check on spectra, which can be plagued by such things as uncertain slopes or slit effects.   For these purposes we require photometry taken in several filters at the same epoch (to within a few days) so that any apparent changes with wavelength are not caused by the variation in overall brightness.  There are several datasets in the literature which contain measurements of IRC +10216 across the 10 $\mu\mbox{m}$ window.  We use these primarily to confirm the normalized shapes of the spectra discussed above, as the large error bars and uncertainties in normalization make epoch to epoch comparisons of the photometry difficult.  In all cases, we used the Vega spectrum of \citet{1995AJ....110..275C} to reduce magnitudes reported in the literature.  We describe our normalization method in detail below.

\subsubsection{Strecker and Ney}
\citet[hereafter SN74]{1974AJ.....79.1410S} observed IRC +10216 at 5 epochs in 1973 at 8.6, 10.7, and 12.2 $\mu\mbox{m}$ (as well as other points outside the N band) from the O'Brien observatory in Minnesota, USA.  Their measurements in Jan, Mar and Apr 1973 provide a nearly contemporaneous check on the spectral shape found by TC74, and provide a useful comparison to the MS76 spectrum.  Estimated errors were reported as $\pm20\%$, which make individual points nearly useless for comparing to spectra.  To overcome this we average the three points from early 1973, and the two points from late 1973, after applying the normalization procedure described below.  We could average all 5, however this method allows for the possibility of short term ($<$ 9 months) variability in the spectral shape of IRC +10216.

\subsubsection{McCarthy, Howell, and Low}
MHL80 reported measurements of IRC +10216's brightness at many epochs in the late 1970s, taken on Kitt Peak, Arizona, using 4, 2.3, 1.5, and 1 m telescopes.  These data were taken in support of their interferometric size measurements.  At only 2 of these epochs (17 Dec., 1977 and 18 Nov., 1978) were measurements made at enough points across the 10 $\mu\mbox{m}$ window to be useful for shape comparisons with our spectra.  We use the MHL80 photometry, with estimated errors of $\pm10\%$, for comparison with the TC74 spectrum.

\subsubsection{The Photometry of Le Bertre}

\citet[hereafter LB97]{1997A&A...324.1059L} obtained bandpass photometry using the European Southern Observatory 1m telescope at La Silla Observatory, Chile, spanning 1985-1988.  These are the same data used in part in the production or our adopted light curve parameters in \citet{1992A&AS...94..377L}. LB97 used filters with central wavelength of 8.38,  9.69, 10.36, and 12.89 $\mu\mbox{m}$ and reported errors of 10\%, 10\%, 10\%, and 15\% respectively.  These data are used here to compare to the IRAS/LRS spectrum.

\subsubsection{TIRCAM}

IRC +10216 was observed in January 1993 by \citet{1996A&A...311..253B} using TIRCAM, a mid-IR camera equipped with a 10x64 array, on the 1.5m Telescopio Italiano Infrarosso at Gornergat (TIRGO), Switzerland.  The filters used had central wavelengths of 8.8, 9.8, 11.7, and 12.5 $\mu\mbox{m}$, with errors of 7\%, 7\%, 15\%, and 15\% reported for IRC +10216.  We compare the photometry of \citet{1996A&A...311..253B} to the UKIRT spectra of MGD98 and the ISO/SWS spectrum.

\subsection{Comparison of Archival Data}
Our main reason for including this archival data is comparison with our new results.  First, though, we can compare the various measurements to each other.  Have the changes our measurements reveal been observed before?  Are these changes part of the regular variability of this object?  The data sets we have extracted from the literature were taken at various points in IRC +10216's two year brightness variations.  To account for this we first normalize the spectra at $\lambda = 10.55\mu\mbox{m}$, averaging across the MIRAC4 $10.55\mu\mbox{m}$ filter bandpass.  This area appears to have had a very stable spectral slope throughout the nearly forty years of observations we consider here.

Normalizing the photometry is a bit more challenging.  Each instrument used had a different photometric system, and authors did not always report results in all filters at each epoch.  Since we are most interested in analyzing the shape of the spectrum for $\lambda > 11\mu\mbox{m}$,  we proceed by first fitting a line to the spectra from $8-11\mu\mbox{m}$ after they were normalized to the MIRAC4 10.55$\mu\mbox{m}$ bandpass.  The spectra all appear to be roughly linear and similar in slope across this region, though with noticeable variation at $\lambda < 9\mu\mbox{m}$.  We then normalize the photometry to this line (which has $F\nu(10.55\mu\mbox{m}) = 1$), using the best-fit normalization factor for each epoch.  We also propagate errors from the fitting procedure to the new normalized photometric points.

\begin{figure}
\begin{center}
\includegraphics[angle=90,scale=.68]{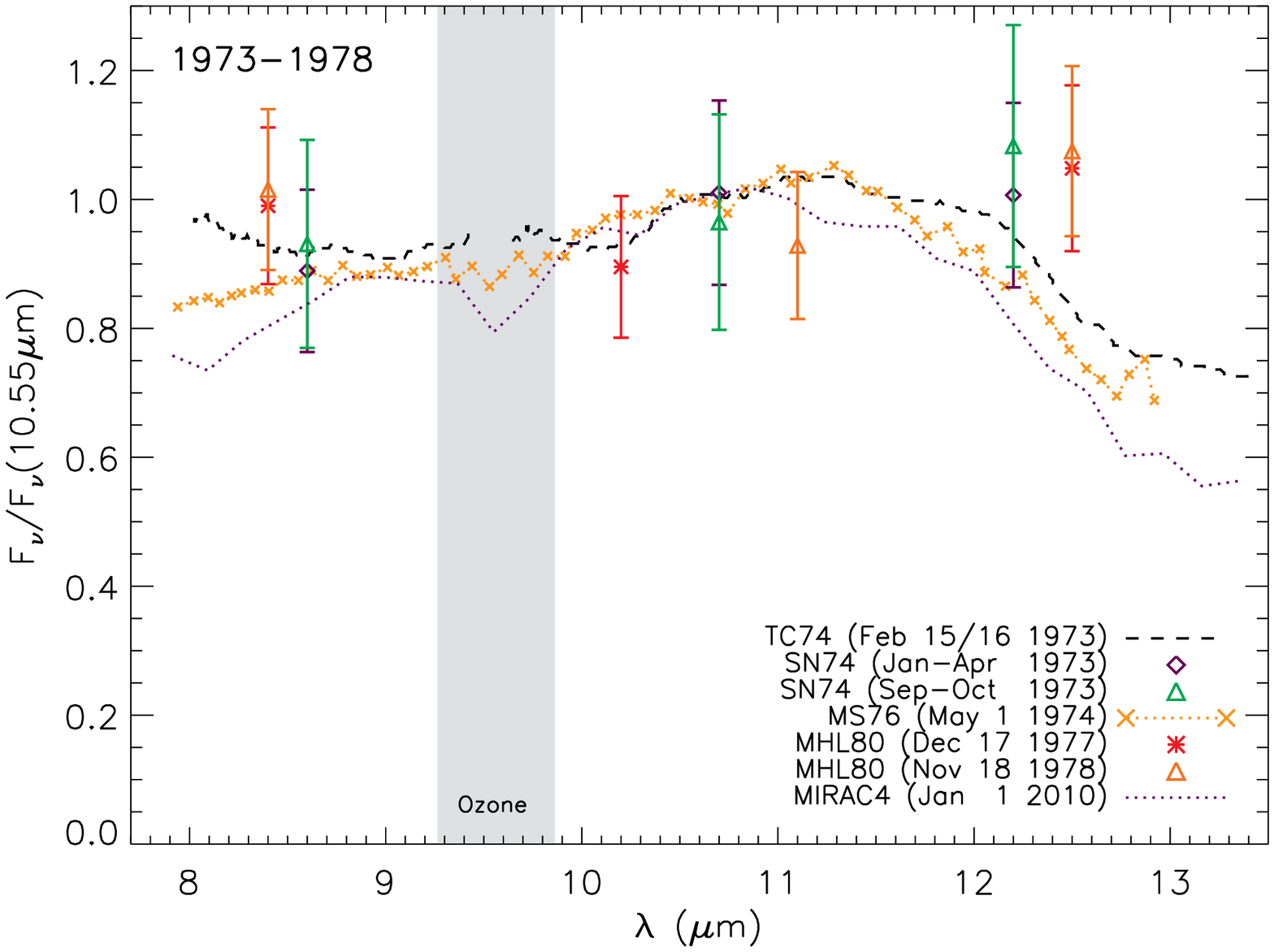}
\end{center}
\caption{IRC +10216 10$\mu\mbox{m}$ spectra and photometry from 1973 to 1978 normalized to $F\nu(10.55\mu\mbox{m}) = 1$.  Here we compare the bandpass photometry of \citet[SN74]{1974AJ.....79.1410S}, the FTS spectrum of \citet[TC74]{1974ApJ...188..545T}, and the CVF photometry of \citet[MS76]{1976PASP...88..294M}, and the bandpass photometry of \citet[MHL80]{1980ApJ...235L..27M}.  We also show the shape of the spectrum in 2010 as reported in this work for reference.  The five SN74 photometry epochs have been averaged (3 points from Jan-Apr, and 2 points in Sep-Oct) to reduce the 20\% uncertainty in the individual points.  Note that the Jan-Apr SN74 data and the TC74 data are essentially contemporaneous.  The photometry generally supports the spectrum shape obtained by TC74, and is also consistent with the MS76 spectrum within 2$\sigma$ uncertainty.
\label{fig:tc74_comp}}
\end{figure}

In Figure \ref{fig:tc74_comp} we show the bandpass photometry of SN74, the FTS spectrum of TC74, the CVF spectrophotometry of MS76, and the photometry of MHL80.  We also show the spectrum of IRC +10216 in 2010 as measured by MIRAC4 in this work.  For $\lambda \lesssim 11\mu\mbox{m}$ the photometry appears to agree nicely with the spectra, but from $12-13\mu\mbox{m}$ it is noticeably brighter in both SN74 and MHL80 - though consistent with the spectra at the $\sim2\sigma$ level.   This could be explained by a slope offset in the TC74 FTS spectrum, however the MS76 CVF spectrum would not likely have such an artifact.  Given the variability of IRC +10216, variations in spectral shape hinted at by this plot might be associated with the 649 day cycle of IRC +10216's luminosity.  We note, however, that the SN74 and the first three TC74 points were taken at nearly identical times.  This would require rapid short term variability in the spectral shape over time scales much shorter than the 649 day brightness variation.

\begin{figure}
\begin{center}
\includegraphics[angle=90,scale=.68]{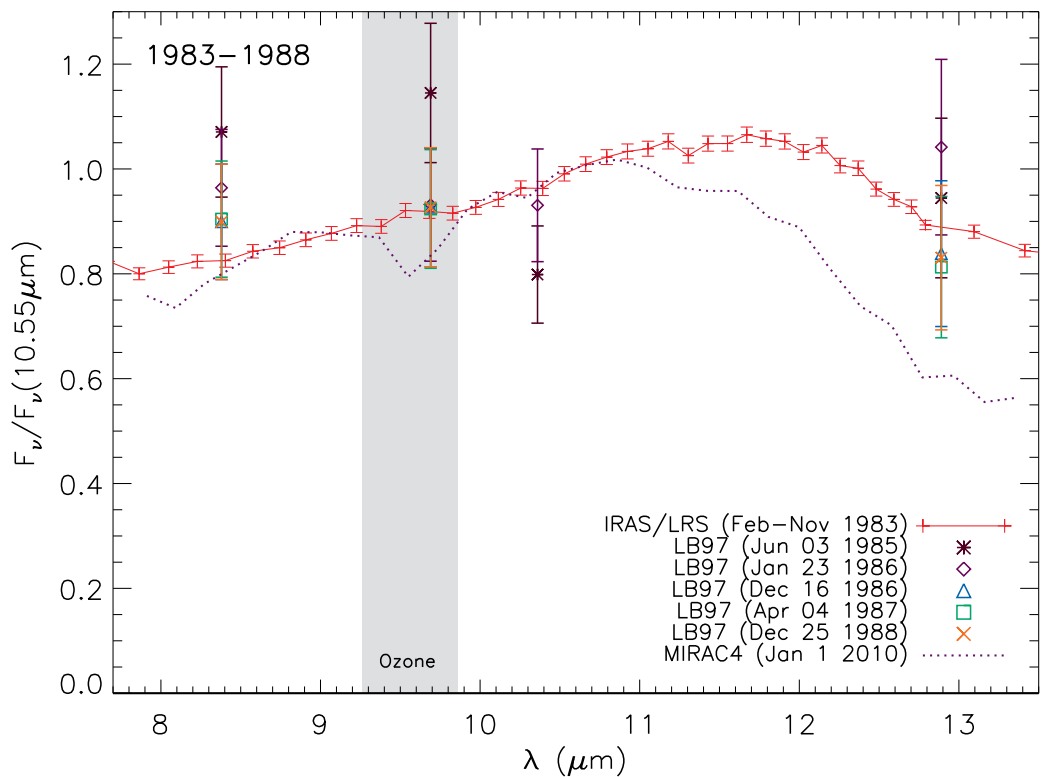}
\end{center}
\caption{IRC +10216 10$\mu\mbox{m}$ spectra from 1983 to 1988.  Here we compare the space-based IRAS/LRS spectrum, and the bandpass photometry of \citet[LB97]{1997A&A...324.1059L}, normalized to $F_\nu(10.55\mu\mbox{m}) = 1$.  The photometry appears to match the IRAS/LRS spectrum well, though as in Fig \ref{fig:tc74_comp} it is consistent with our 2010 data at the 2$\sigma$ level.  The LB97 points are taken far enough apart in time that we do not average in case there is short term variation in the shape.  
\label{fig:iras_comp}}
\end{figure}

A comparison of data from the 1980s is provided in Figure \ref{fig:iras_comp}.  The photometry of LB97 and the IRAS/LRS spectrum agree well within the $1\sigma$ error, with the exception of the June 1985 point which appears to have been strongly affected by atmospheric $O_3$.  As in Figure \ref{fig:tc74_comp} we see that the photometry is generally consistent with the 2010 shape (and that measured by MS76) within the $2\sigma$ uncertainty.

\begin{figure}
\begin{center}
\includegraphics[angle=90,scale=.68]{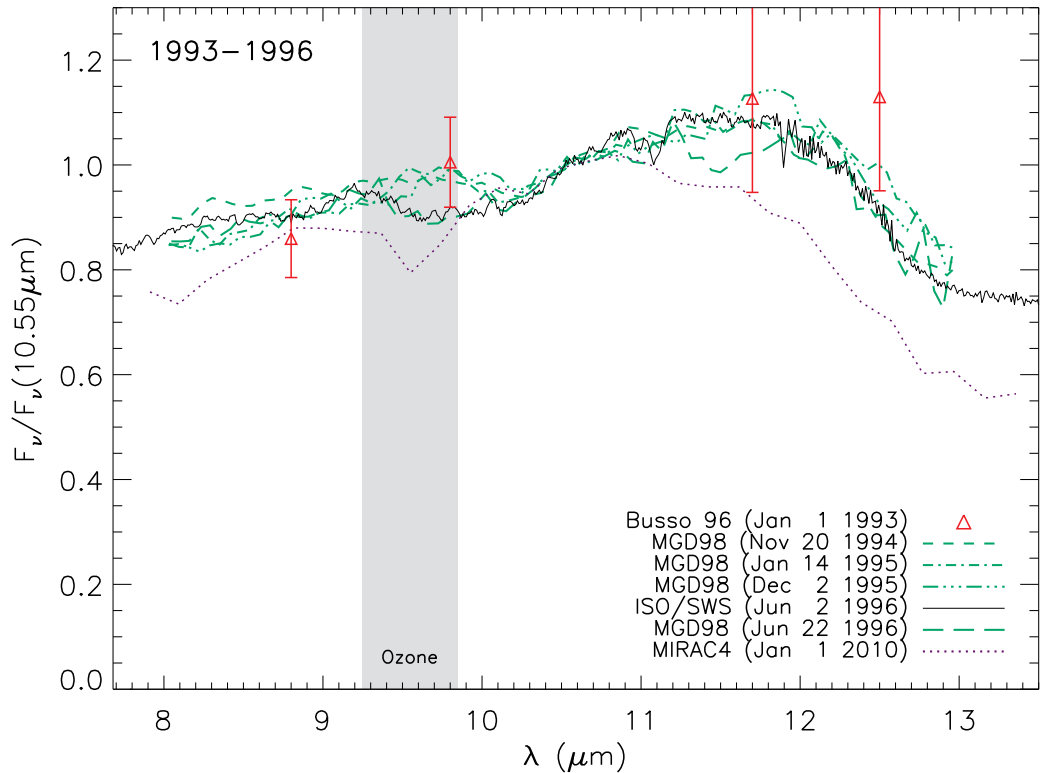}
\end{center}
\caption{IRC +10216 10$\mu\mbox{m}$ spectra from 1993 to 1996.  Here we compare the bandpass photometry of \citet{1996A&A...311..253B}, the UKIRT spectra of \citet[MGD98]{1998ApJ...502..833M}, and the space-based spectrum obtained by ISO/SWS.  We normalized the data to $F_\nu(10.55\mu\mbox{m}) = 1$. All three data sets are in good agreement during this period, which spans three and a half years  and well samples nearly two luminosity periods. The comparison with our 2010 data clearly shows that the change in the spectrum at $\lambda > 11 \mu\mbox{m}$ is not simply associated with the regular 649 day variation in brightness of IRC +10216.
\label{fig:busso_comp}}
\end{figure}

We continue our decade by decade comparisons with Figure \ref{fig:busso_comp}, which shows data from the 1990s.  In this case the data agree quite well across the entire $10\mu\mbox{m}$ window, and we note especially the agreement between the UKIRT and ISO/SWS spectra taken twenty days apart in June 1996.  These measurements span three and a  half years, and two full 649 day periods, so unlike the previous decades we can say with some confidence that there is no variation in shape, large enough to explain our 2009/2010 results at $\lambda \gtrsim 11 \mu\mbox{m}$, occurring as part of the regular 649 day variability of IRC +10216 during this time period.

\section{Discussion}
\label{sect:discussion}
\subsection{Changes in the 10 $\mu\mbox{m}$ spectrum of IRC +10216}
Figure \ref{fig:flux_off} shows a comparison of nearly four decades of N-band spectra of IRC +10216 with the data normalized to $F_{\nu}(10.55\mu\mbox{m}) = 1$.  We also include our 2009 photometry, which matches our 2010 grism spectrum very well.  It is clear that a significant change at wavelengths longer than $\sim11\mu\mbox{m}$ was recorded in our 2009 and 2010 data when compared to the mid-1990s, and that a similar shape was observed in the early 1970s.  The negative slope of the spectrum has become steeper and the continuum is lower at wavelengths redder than $13\mu\mbox{m}$.  The close match between our bandpass photometry and grism spectrum gives confidence that this is not merely a calibration error, and since they are taken 1 year apart at different luminosity phases the shape appears to be stable at the time of our observations.

In Figure \ref{fig:time_series} we plot the flux ratio $F_{\nu}(12.5\mu\mbox{m})/F_{\nu}(10.55\mu\mbox{m})$ vs. time, where flux at 12.5 $\mu$m was calculated as the mean between 12 and 13 $\mu$m for each of the spectra.  This figure illustrates the change in the shape of the spectrum over time.  The change from 1996 to 2009 does not occur as part of the 649 day Mira variability, as evidenced by the mid 1990s data. We discuss two possible interpretations of this plot further below.
 
\citet{1999ApJ...521..261M} found long term changes in three carbon stars, including IRC +10216 which had the smallest change.  They used the MS76, IRAS/LRS, and MGD98 data, and reported a change in spectral slope across the 10 $\mu\mbox{m}$ window ($8-13\mu\mbox{m}$) from the early 1970s to 1996.  We now see a change in the opposite direction from 1996 to 2009 in IRC +10216's spectrum.

Perhaps the simplest interpretation of these results is that we have recorded an episode of irregular variability (i.e. recurring but not periodic changes) in the spectrum, rather than a trend.  Due to the sparse sampling and relatively long time periods between measurements (e.g. the gap from 1978 to 1983, or from 1996 to 2009) we can make no statements about how often this irregular variability occurs.  In the case of the 2009-2010 MIRAC4 data this new shape lasts for at least one year, or half the period.  Given the clumpy structure of the CSE \citep{1998A&A...333L..51W}, the anisotropic nature of the mass loss history \citep{2006A&A...455..187L}, and the rapid variations seen in the inner regions of the CSE \citep{1998A&A...334L...5H, 2000ApJ...543..284T}, relatively rapid and irregular variability in IRC +10216's spectrum might be expected.

We also consider the possibility that these changes are occurring over a longer term.  Though sparsely sampled in time, figure \ref{fig:time_series} has the appearance of a smooth change over $\sim40$ years.  Other circumstantial evidence for the longer term variability is provided in the findings of previous studies we discussed in section \ref{sect:spatial_domain}.  \citet{2000A&A...359..707M} found evidence for $\sim 40$ year modulation in the expanding dust shells around IRC +10216 from deep V band observations.  \citet{2002A&A...392..921M} claimed that the current episode of dust production began roughly in the 1950s, and that the mass loss rate had recently increased.  \citet{2000ApJ...543..861M} claimed that dust production had stopped by the end of the 1990s in contradiction to \citet{2002A&A...392..921M}.  We won't attempt to resolve these contradictory modeling results, but rather take them as evidence that something had changed in the mass loss rate at the end of the 1990s.  This idea, coupled with the correspondence between the periods evident in the V band observations, the hypothesized start date of the current mass loss episode, and the timing of the changes evident in the spectrum by 2010, supports the possibility of a longer term change in IRC +10216.

In any case, we can place the shape of IRC +10216's N band spectrum in context with other carbon stars using the ``Carbon-Rich Dust Sequence'' of \citet*{1998AJ....115..809S} (hereafter SLMP98).  Their system is based on 96 carbon-rich variable stars observed by IRAS/LRS, from which they subtract a 2400K blackbody to remove the stellar continuum.  The blackbody is fit to the wavelength range 7.67-8.05$\mu\mbox{m}$.  After subtraction the spectra were normalized, and then were  grouped by inspection according to the shape of the $\sim11\mu\mbox{m}$ SiC feature and the presence and strength of the $9\mu\mbox{m}$ feature (which we acknowledge is likely not real).  In Figure \ref{fig:sloan_class} we show their sequence, formed from averaging and smoothing each spectrum in the class, as the solid black curves.  As we discussed in Section \ref{sect:intro_sic}, in light of the results of \cite{2006ApJ...652.1654T} we do not treat this sequence as reflecting the evolutionary state of these carbon stars.  Nevertheless, we have found the system of SLMP98 to be a useful atlas of the SiC spectra in carbon stars and as an aid to interpreting our results.

In Figure \ref{fig:sloan_class}, we show the IRAS/LRS spectrum of IRC +10216 in red, which was classified by SLMP98 as ``Red'', and cited as the prototype of that class.  We also show our 2010 spectrum in blue, continuum subtracted and normalized according to the above prescription.  In this framework, it appears that the center of the SiC peak has shifted blue-ward, and the 9$\mu\mbox{m}$ region is enhanced, which appears to be true regardless of whether the minimum at $\sim9.5\mu\mbox{m}$ is caused by ozone (it almost certainly is).  In 2010, IRC +10216 is a better match for the ``Broad 2'' (Br2) class in the SLMP98 system.  Even though SLMP98 does not represent an astrophysical sequence for carbon stars, we see that the shape we observed does occur in other carbon stars.

Regardless of whether this change occurs irregularly on timescales of a few years, or represents multi-decade variability,  such variability could easily be overlooked in similar Carbon stars.  The SLMP98 system is especially useful here, as it provides a catalog of spectra and targets for follow-up observations to test this possibility.  We find that the IRAS/LRS spectra of sources RV Cen (Br1) and CR Gem (Br2) are good qualitative matches for our 2010 MIRAC4 spectrum.  We also checked the SiC feature of LP And, the one other source in the Red classification that was also observed by ISO/SWS, and found that it does not appear to have changed between the two observations.  All of the ``Red'' and ``Broad'' sources deserve future observations in this wavelength range to check for any irregular variability or long-term changes.

In addition to follow up observations of IRC +10216 and similar carbon stars, fully understanding the changes reported here will require detailed modeling of the dusty CSE of IRC +10216. Given the wealth of data on this object and its complex asymmetric structure, such modeling is beyond the scope of the current paper (see \citet{2001A&A...368..497M} for an example of such a comprehensive effort).

\subsection{The spatial signature of SiC emission}
In Figure \ref{fig:spectra_fwhm} we place our 2010 spectrum and corresponding FWHM measurement on a common wavelength axis.  Here we have ``deconvolved'' IRC +10216 by subtracting the FWHM of the PSF in quadrature, in order to estimate its intrinsic size.  IRC +10216 exhibits an apparent increase in size, quite separate from the effect of diffraction alone, in the wavelength range of the SiC emission feature.  Fully understanding the spatial signature of the spectral emission feature will require additional observations at different luminosity phases, including 2D imaging, as well as detailed modeling.

A likely interpretation is that we have observed the effect of radiative transfer through the CSE.  SiC has higher opacity at the wavelengths of the feature, hence we observe photons produced further from the star at those wavelengths, which causes an apparent increase in size in the feature's part of the spectrum.  The change in apparent size can be thought of as mapping the optical depth of the CSE as a function of wavelength.

We can use these results to establish the plausibility of the changes in the spectrum being caused by the outflow of material from the star.  If we treat the estimated intrinsic size at $13 \mu\mbox{m}$, $\sim 70$AU, as the diameter of the CSE, and take the time for the changes in the CSE to occur as 12.5 years (1996.5 to 2009), we find an estimated outflow velocity of 13 $\mbox{km sec}^{-1}$.  Estimates in the literature for the expansion velocity of IRC +10216's CSE range from 12 to 17 $\mbox{km sec}^{-1}$ \citep[and references therein]{2001A&A...368..497M}.  Using their model, \citet{2001A&A...368..497M} calculated the deprojected radial velocities of the clumps observed by \citet{2000A&A...357..169O} as $\sim 15 \mbox{km sec}^{-1}$.  Hence, we find that the changes in the spectrum from 1996 to 2009 can plausibly be explained by evolution of the CSE, and the resultant estimated outflow velocity is in good agreement with previous estimates.

A comment on the possible asymmetry evident in Figure \ref{fig:fwhm} should be made.  The FWHM is smaller in the East-West direction than in the North-South direction, roughly indicating an elongation towards the N-NE, exactly as expected from imaging studies at shorter wavelengths.  We are cautious with this result, however.  An important consequence of the profiles of IRC +10216 following a Lorentzian is that small offsets in the slit will cause changes in the apparent shape of the object.  Whereas when a 2D Gaussian is sliced somewhere off the peak the same parameters (i.e. FWHM) describe the resulting curve, the same is not true for a Lorentzian.  Since we were not able to repeat the observations at each position angle due to time limitations, we have no way to estimate the uncertainties in the individual FWHM curves at different position angles due to slit alignment.

\section{Conclusion}
We have presented new photometric and spectroscopic measurements of the well-studied carbon star IRC +10216 in the N band atmospheric window.  When compared to nearly 4 decades of prior observation we find that a significant change (decrease in brightness) appears to have occurred in the $11-13.5 \mu\mbox{m}$ region of the spectrum, which includes the SiC emission feature, between 1996 and 2009.  Measurements taken in early 1970s appear to match the 2009/2010 shape, but data from the 1980s and 1990s does not.  We discussed two possible explanations for these changes.  We may have observed an episode of irregular variability distinct from IRC +10216's regular $\sim2$ year Mira variability.  We also consider it a possibility that we have observed a long term change occurring over several decades.

Critical to our reduction of the grism spectrum was the stable, high Strehl, diffraction limited information provided by the MMTAO system, which was needed to correct for the differential slit loss between the extended source IRC +10216 and the point source standard.  This spatial information, which allows us to analyze size vs. wavelength, shows that the SiC emission feature has a clear spatial signature in the dust surrounding IRC +10216.  The CSE exhibits an increase in apparent size of $\sim30\%$ between $10.2$ and $11.6\mu\mbox{m}$ compared to the continuum on either side of the SiC feature.  This is likely tracing the higher optical depth due to SiC in the $\sim 70$ AU  CSE.  We used this estimate of the object's intrinsic size to establish that the observed spectrum change over 12.5 years can plausibly be associated with the evolution of the dusty CSE of IRC +10216 given 12-17 km s$^{-1}$ outflow velocities.

\acknowledgements
We thank the anonymous referee for a thorough review and for providing insight into 1970s spectrophotometry, which resulted in a much improved manuscript.  We thank John Monnier for providing data from UKIRT in tabular form.  We thank John Bieging for his comments on a draft of this manuscript.  JRM is grateful for the generous support of the Phoenix ARCS Foundation.  AJS acknowledges the generous support of the NASA GSRP program.  This work and LMC were supported by the NASA Origins program, and the NSF AAG and TSIP programs.

Facilities:\facility{MMT}

\clearpage
\appendix
\section{The possibly erroneous 8.7$\mu\mbox{m}$ photometry from 2009}

\begin{landscape}
\begin{figure}
\begin{center}
\includegraphics[angle=90,scale=.525]{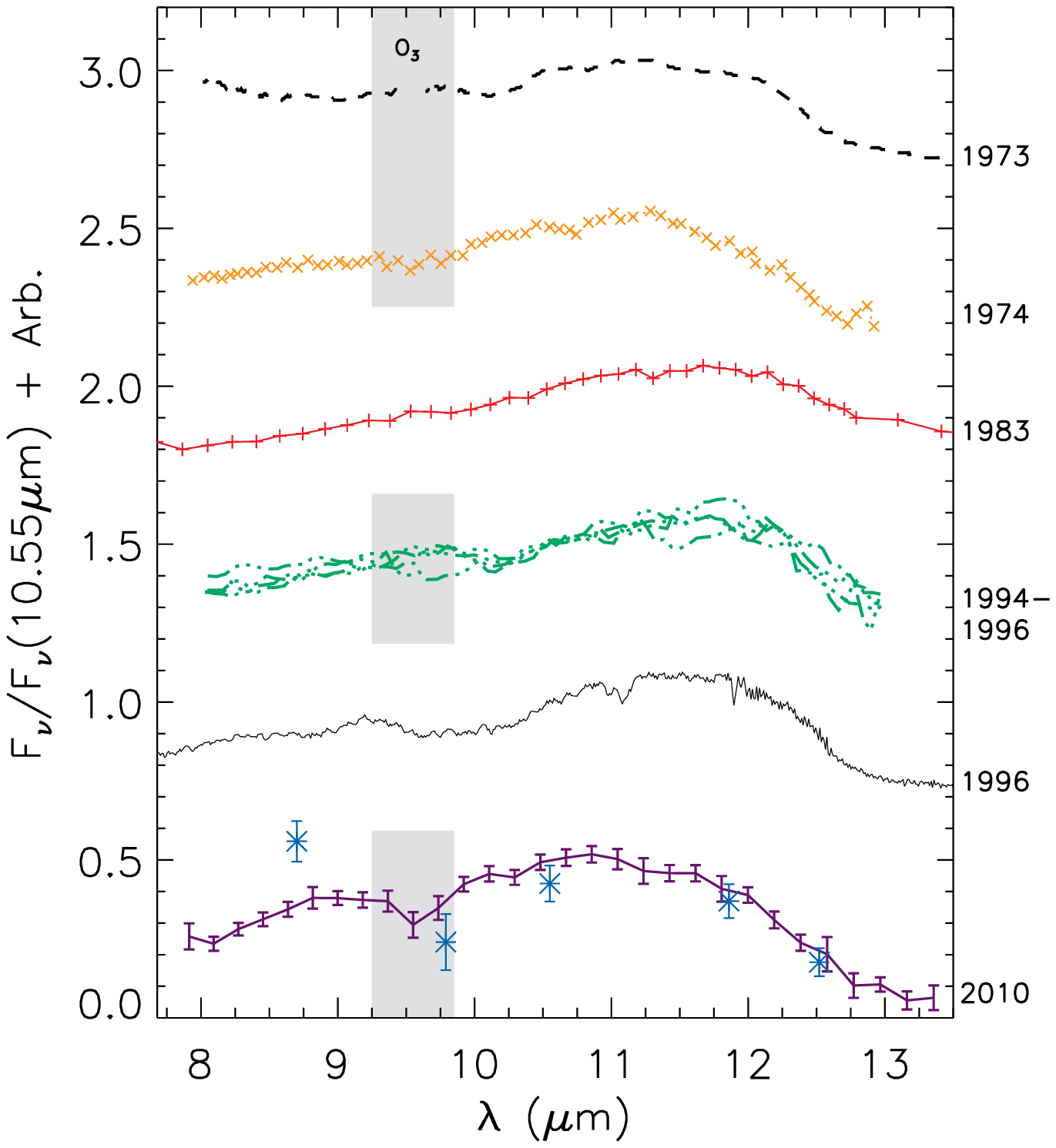}
\includegraphics[angle=90,scale=.525]{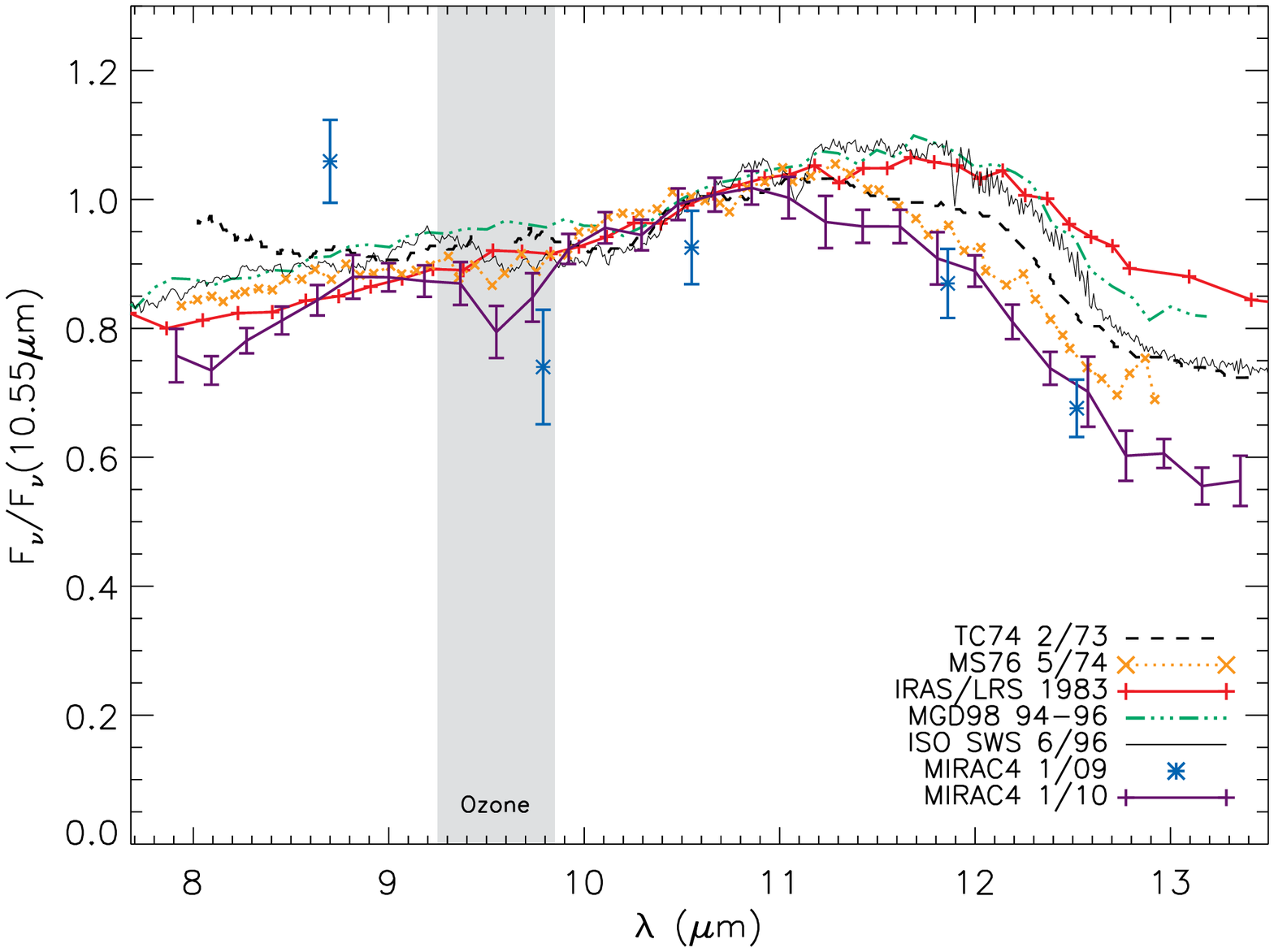}
\end{center}
\caption{Our new N band spectrum and photometry compared to previous observations of IRC +10216 spanning nearly 4 decades.  As in Figures \ref{fig:tc74_comp}-\ref{fig:busso_comp} the data have been normalized at $10.55\mu\mbox{m}$.   At left we have added an arbitrary constant to offset each epoch.  The MIRAC4 photometry and grism spectrum, taken a year apart, match very well from $9.8\mu\mbox{m}$ to $12.5\mu\mbox{m}$.   We present the same data at right without the offset.  Of all the data from prior epochs,  the spectrophotometry of MS76 is most similar to the 2009/2010 MIRAC4 data red-ward of $11\mu\mbox{m}$.  The archival data rule out these changes being simply related to the regular 649 day Mira variability exhibited by IRC +10216.
\label{fig:flux_off}}
\end{figure}
\end{landscape}

\begin{figure}
\begin{center}
\includegraphics[angle=90,scale=.68]{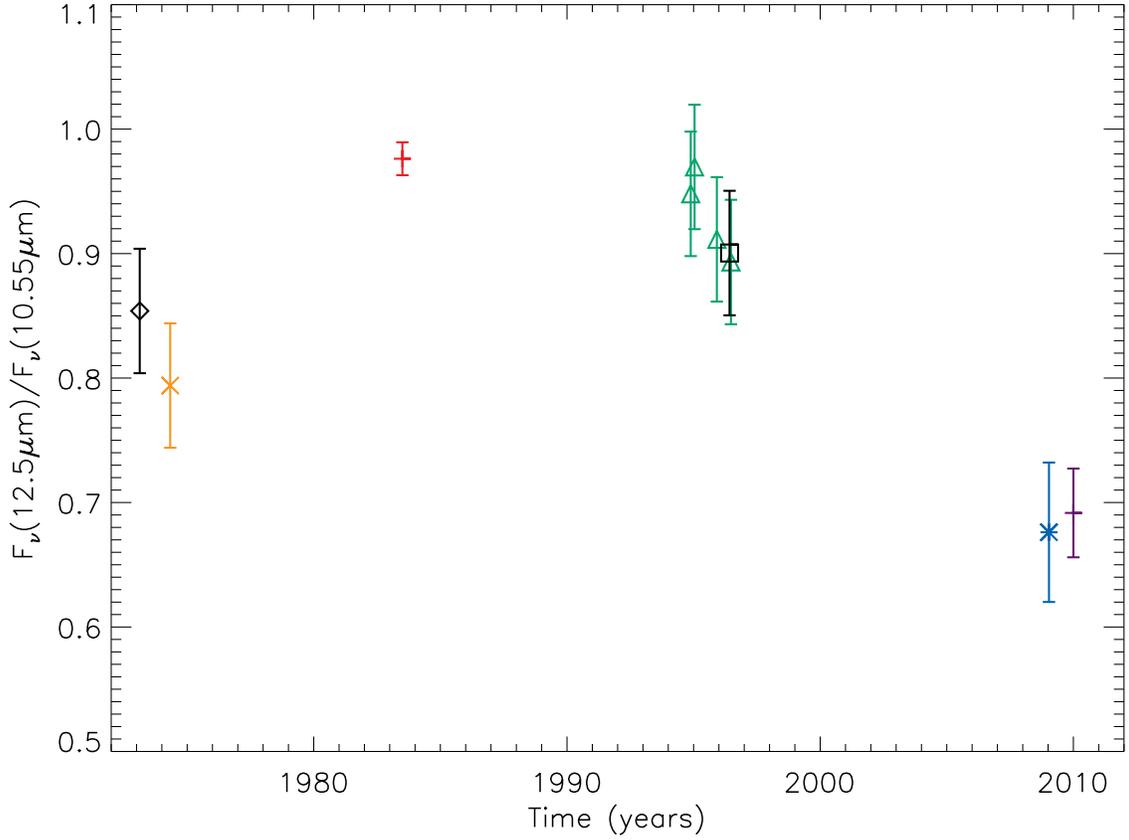}
\end{center}
\caption{The flux ratio $F_{\nu}(12.5\mu\mbox{m})/F_{\nu}(10.55\mu\mbox{m})$ vs. time.  Colors are the same as in Figure \ref{fig:flux_off}.  Flux at 12.5 $\mu$m was calculated as the mean between 12 and 13 $\mu$m.    MIRAC4 and IRAS/LRS errors are as given.  We adopt local or relative error of $\pm5\%$ for the other data sets where such errors were not given.  This plot illustrates the change in the shape of the spectrum over time, highlighting variability not associated with the regular 649 day Mira luminosity variations of IRC +10216.  Whether this is a recurring spectrum shape which occurs at irregular intervals, or a longer term ($\gtrsim 40$ year) periodicity cannot be determined from the available data.
\label{fig:time_series}}
\end{figure}

\begin{figure}
\begin{center}
\includegraphics[angle=90,scale=.8]{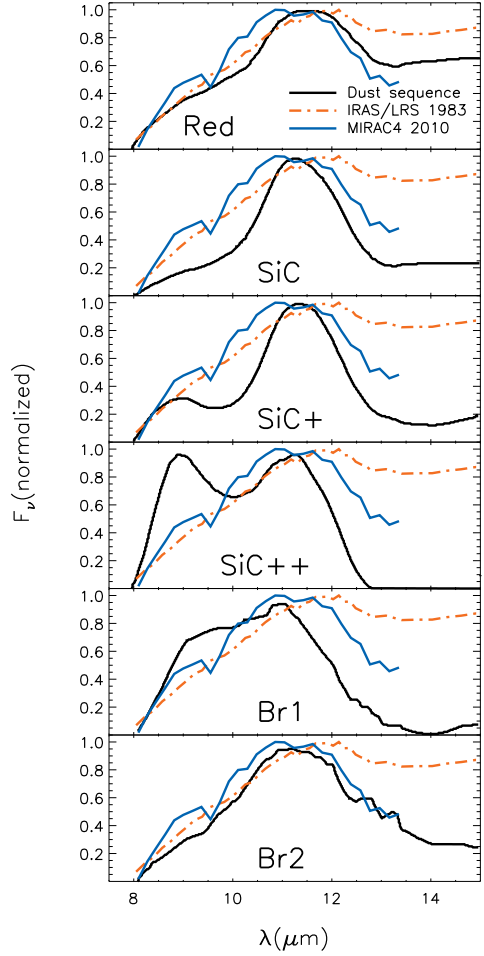}
\end{center}
\caption{The current state of IRC +10216 plotted on the ``Carbon-Rich Dust Sequence'' classification system of SLMP98.  This system was based on the IRAS/LRS spectra of 96 carbon-rich AGB stars, and involves subtracting a 2400K blackbody (an approximation for the stellar continuum), normalizing, and visually inspecting the resulting curves.  The heavy black curves are the summed and smooth spectra used to illustrate the sequence, and we show the 1983 IRAS/LRS spectrum (dotted red) and our 2010 MIRAC4 spectrum (blue).  SLMP98 cited IRC +10216 as the prototype of the Red class, but it now (2010) appears to be a better match to the Broad 2 (Br2) spectra for $\lambda > 11\mu\mbox{m}$.  Though the SLMP98 system does not represent an astrophysical sequence for C stars, it is useful in this case to show that our measurement of IRC +10216's spectrum matches other C stars.
\label{fig:sloan_class}}
\end{figure}

\begin{figure}
\begin{center}
\includegraphics[angle=90,scale=.68]{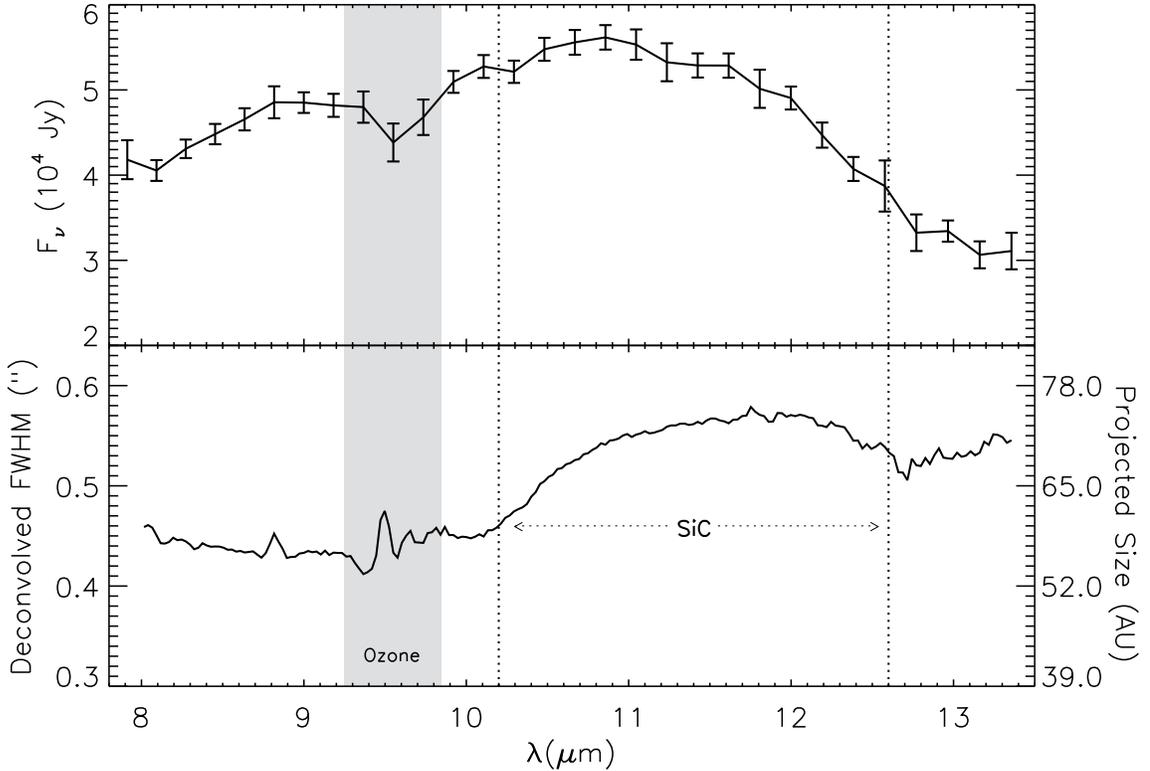}
\end{center}
\caption{The spectral and spatial signatures of SiC dust around IRC +10216.  The well known SiC spectral feature can be seen in the MIRAC4 grism spectrum from 2010 in the top panel.  In the bottom panel we have deconvolved the FWHM of IRC +10216 by subtracting the PSF FWHM in quadrature in order to estimate its intrinsic size, after averaging the four PAs.  To provide a physical scale, we follow \citet{2001A&A...368..497M} and adopt 130pc for the distance to IRC +10216 and calculate the projected size corresponding to the FWHM. An increase in the size of IRC +10216, clearly corresponding to the SiC feature, is evident (we have used the typical bounds for this feature as found by \citet{2003ApJ...594..642C}).
\label{fig:spectra_fwhm}}
\end{figure}

Here we further discuss the MIRAC4 2009 $8.7\mu\mbox{m}$ photometry data point, which appears significantly over-luminous in Figure \ref{fig:flux_off}.  We see no evidence that a change in weather or seeing affected the 8.7$\mu\mbox{m}$ PSF measurement without affecting the others, and there has so far been no evidence that this filter has a leak during other BLINC/MIRAC4 observations.  Nevertheless, since this data point was not confirmed by our follow-up grism data and is in a high-flux regime of the detector not well understood, nor tested by any of our other data, we remain suspicious of the 8.7$\mu\mbox{m}$ photometry.

The 8.7$\mu\mbox{m}$ filter had $\sim85\%$ higher peak counts than the next brightest 10.55$\mu\mbox{m}$ filter (both are above the background), due in part to its width ($\sim40\%$ wider than $10.55\mu\mbox{m}$), as well as differences in detector quantum efficiency.  This led us to suspect that the most likely culprit for the discrepancy would be non-linearity of the MIRAC4 detector, which does exhibit an increase in slope at higher fluxes.  As part of the normal preparation for observing a linearity measurement was performed in a laboratory at Steward Observatory one week prior to these observations.  Unfortunately the bias level appears to have changed in the intervening period, which prevents us from directly applying the curve to our data.  Ordinarily this is of no consequence when using chop and nod background subtraction, so it was not noticed until long after the observations were complete.  We can still perform a worst case analysis though, and decide what effect, if any, non-linearity has on the $8.7\mu\mbox{m}$ measurement.

\begin{figure}
\begin{center}
\includegraphics[scale=.68]{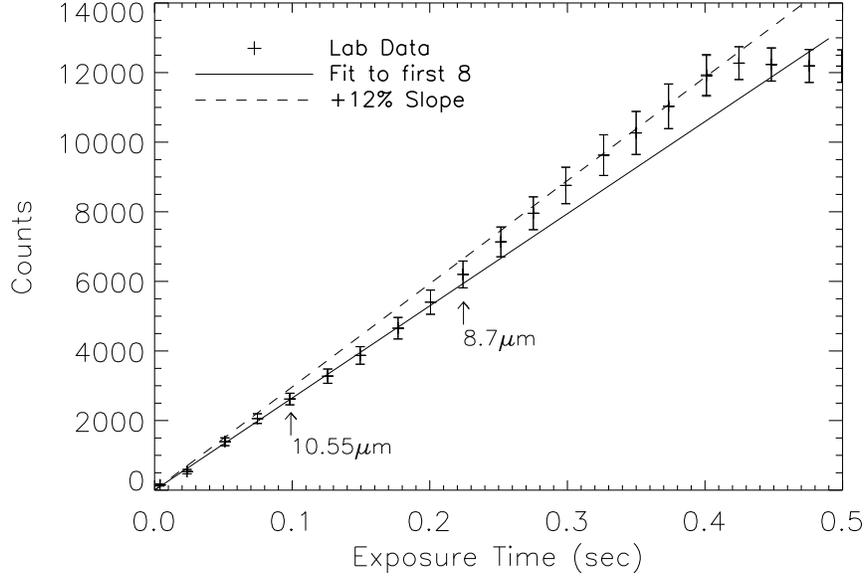}
\end{center}
\caption{MIRAC4 detector linearity measurement.  The data were taken prior to the 2009 observations in a laboratory, and show that the detector becomes non-linear at higher fluxes, exhibiting an increase in slope which could be an explanation for the high flux detected at $8.7\mu\mbox{m}$ (see Figure \ref{fig:flux_off}).  We don't apply the curve directly to the data due to an unnoticed change in the detector bias that occurred between this measurement and the observations.  The solid line is the fit to the first 8 data points.  This fit was used to bias subtract the data. The dashed line has a 12\% higher slope, chosen to illustrate a worst case scenario where every pixel in the $8.7\mu\mbox{m}$ that is brighter than the peak in the $10.55\mu\mbox{m}$ data has that slope.  We also note the actual peak counts per read in the 8.7$\mu\mbox{m}$ filter prior to background subtraction, and the peak counts per read in the 10.55$\mu\mbox{m}$ after background subtraction (where the arrows intersect the solid line).  The peak pixel may have become slightly non-linear, but the integrated non-linearity effect was likely $< 1\%$, causing us to rule it out as an explanation for the high flux at $8.7\mu\mbox{m}$. 
\label{fig:linearity}}
\end{figure}

We start by assuming that the peak counts value in the $10.55\mu\mbox{m}$ image is the last linear value and that all pixels in the $8.7\mu\mbox{m}$ data above this value have a different slope.  With this definition the fraction of flux in non-linear pixels in the $8.7\mu\mbox{m}$ image is $F_{NL} = 32\%$.  We can then estimate the change in slope $\Delta L$ required to produce the change in total flux: $\Delta L = \frac{\Delta F}{F_{NL}}$.  Since we are trying to explain a discrepancy of $\Delta F = 30\%$ we need a slope change of nearly $100\%$.  Figure \ref{fig:linearity} shows the laboratory linearity measurement, along with a fit to the lower portion of the curve.  The data have been bias subtracted using the fit.  We also show a line with 12\% higher slope, which represents the worst case prior to saturation.  The arrows on the plot indicate the peak counts per read in the 8.7 filter and in the 10.55 filter, where the 8.7 point is from raw counts prior to background subtraction (and so includes the unknown bias level) and the 10.55 point is background subtracted.  Figure \ref{fig:linearity} demonstrates that even in the worst case scenario where every non-linear pixel has a $12\%$ higher slope, non-linearity can explain at most $\sim4\%$ of the excess flux in the $8.7\mu\mbox{m}$ filter. We see that the non-linearity in fact likely causes a less than $1\%$ error.  Given this result, we have no reason to reject the $8.7\mu\mbox{m}$ data point out of hand due to non-linearity.

Finally, we note that the archival photometry presented in Section \ref{sect:archive} has several examples of apparent excesses at $\sim8.7\mu\mbox{m}$, but such a feature never appears in the spectra.  This points to unquantified systematics in broadband photometry in this region, which is bounded closely by variable water vapor and ozone.  We have tried applying a correction for differences in spectral shape between object and standard, and assuming large changes in PWV between object and standard using the ATRAN model, and so far have not found an explanation for these excesses.

At this point in time, the evidence is inconclusive and we remain suspicious of the 2009 MIRAC4 $8.7\mu\mbox{m}$ filter photometry.  The change from 2009 to 2010 would require a significant decoupling at this wavelength from the rest of the spectrum with regards to the regular 649 day variability.  Though we have ruled out non-linearity as a cause, the per-pixel flux achieved is the highest ever observed with MIRAC4 and we can't yet rule out changes in the read-out artifacts (e.g. cross-talk) at higher flux.  Further observations, with both bandpass photometry and grism data taken at the same epoch, are required to fully understand our 2009 data point.

\clearpage
\bibliographystyle{apj}
\bibliography{males}

\clearpage

\clearpage

\clearpage

\clearpage

\clearpage

\clearpage

\clearpage

\clearpage

\clearpage

\clearpage

\clearpage

\clearpage

\clearpage

\begin{deluxetable}{lcccccc}
\tablewidth{418pt}
\tablecaption{Observations of IRC +10216 and standards.\label{tab:obslog}}
\startdata
\hline
\hline
Object & Filter  & Airmass & Pos.  & AO Speed &  No. & Total Exp.\\
       & ($\mu\mbox{m}$)\tablenotemark{1} &         & Angle\tablenotemark{2} & (Hz)     &  Frames \tablenotemark{3} & Time (sec)\tablenotemark{3}\\
\hline
\cutinhead{2009 13 Jan UT}
\hline
IRC +10216 & $8.7$  & 1.35 & N/A & Off & 8 & 1.76\\
IRC +10216 & $9.79$  & 1.38 & N/A & Off & 16 & 3.52\\
IRC +10216 & $10.55$  & 1.32 & N/A & Off & 12 & 2.64\\
IRC +10216 & $11.86$  & 1.40 & N/A & Off &  4 & 0.88\\
IRC +10216 & $12.52$  & 1.31 & N/A & Off &  12 & 2.64\\
$\mu$ UMa & $8.7$ & 1.23 & N/A & Off &  24 & 5.28\\
$\mu$ UMa & $9.79$ & 1.22 & N/A & Off &  20 & 4.40\\
$\mu$ UMa & $10.55$ & 1.18 & N/A & Off &  20 & 4.40\\
$\mu$ UMa & $11.86$ & 1.20 & N/A & Off &  24 & 5.28\\
$\mu$ UMa & $12.52$ & 1.19 & N/A & Off &  16 & 3.52\\
\tableline
\cutinhead{2010 1 Jan UT}
\tableline
IRC +10216 & Grism & 1.06 & 107.0 & 25 &  5 & 0.040\\
IRC +10216 & Grism & 1.10 & 129.4 & 25 &  2 & 0.016\\
IRC +10216 & Grism & 1.12 & 2.6 & 25 &  8 & 0.064\\
IRC +10216 & Grism & 1.25 & 32.6 & 25 &  6 & 0.032\\
$\mu$ UMa & Grism & 1.02 & 73.1 & 100 &  6 & 10.000\\
\tableline
\cutinhead{2010 2 Jan UT\tablenotemark{4}}
\tableline
$\beta$ Gem & Grism & 1.01 & 144.3 & 150 &  20 & 200.0\\
$\beta$ Gem & Grism & 1.07 & 164.0 & 550 &  8 & 80.0\\
$\mu$ UMa & Grism & 1.03 & 133.7 & 550 &  8 & 80.0\\
$\mu$ UMa & Grism & 1.06 & 92.1 & 550 &  8 & 80.0\\
\tableline
\enddata
\tablenotetext{1}{Filter widths are given in Table \ref{tab:phot}.}
\tablenotetext{2}{Position angle of the slit.}
\tablenotetext{3}{After rejecting bad chops, frames with excessive pattern noise, and bad slit alignment.}
\tablenotetext{4}{Data from this night were only used to check our slit loss correction procedure.}
\end{deluxetable}

\clearpage

\begin{deluxetable}{lccccccc}
\tablewidth{0pt}
\tablecaption{Bandpass photometry of IRC +10216 from 13 Jan 2009 UT.\label{tab:phot}}
\startdata
\hline
\hline
Filter  & Width \tablenotemark{1} &  $F_{\nu}$ &  Obj\tablenotemark{2} $\sigma$  & PSF\tablenotemark{3} $\sigma$  & Std.\tablenotemark{4} $\sigma$ & Atm.\tablenotemark{5} $\sigma$ & Total $\sigma$ \\
$(\mu\mbox{m})$  & $(\mu\mbox{m})$       &    (Jy)    &    (\%)        &    (\%)       &    (\%)       &   (\%)        &   (\%) \\
\hline
\hline
8.7     & (8.08-9.32)     &  45099     &    1.63        &  0.78        &      2.4       &  4            &  5.0\\
9.79    & (9.33 - 10.25)  &  31514     &    2.03        &  1.50        &      2.4       &  11           &  11.5\\
10.55   & (10.06 - 11.04) &  39408     &    1.50        &  1.52        &      2.4       &  4            &  5.1\\
11.86   & (11.29 - 12.43) &  37035     &    1.65        &  1.37        &      2.4       & 4             &  5.1\\
12.52   & (11.94 - 13.11) &  28790     &    1.50        &  2.10        &      2.9       & 4             & 5.6\\
\tableline
\enddata
\tablenotetext{1} {Half power points of the manufacturer provided curves.}
\tablenotetext{2} {The measurement uncertainty in the IRC +10216 photometry, estimated empirically.}
\tablenotetext{3} {The measurement uncertainty in the $\mu$ UMa photometry, estimated empirically.}
\tablenotetext{4} {Mean value of the total uncertainty given by \citet{1996AJ....112.2274C} between the half power points.}
\tablenotetext{5} {Based on the global and local telluric uncertainties of \citet{2010ApJ...711.1280S} from the following night.}
\end{deluxetable}

\clearpage

\begin{deluxetable}{lccccc}
\tablewidth{400pt}
\tablecaption{Grism photometry from 1 Jan 2010 UT.\label{tab:grismspect}}
\startdata
\hline
\hline
$\lambda$ & $F_{\nu}$ & Meas.\tablenotemark{1} $\sigma$ &  Local\tablenotemark{2} $\sigma$  & Global\tablenotemark{3} $\sigma$ & Total\tablenotemark{4} $\sigma$\\
 ($\mu\mbox{m})$  &  (Jy)     &     (\%)       &     (\%)             &  (\%)           &   (\%)\\
\hline
\hline
7.913  & 41810.8  &   4.94   &  5.46   &  2.7   &  8.24 \\
8.092  & 40541.0  &   1.79   &  3.01   &  2.7   &  6.84 \\
8.273  & 43087.7  &   0.71   &  2.52   &  2.7   &  6.64 \\
8.453  & 44816.2  &   1.12   &  2.66   &  2.7   &  6.69 \\
8.635  & 46551.8  &   1.16   &  2.80   &  2.7   &  6.70 \\
8.817  & 48551.0  &   3.03   &  3.86   &  2.7   &  7.26 \\
8.999  & 48513.1  &   0.80   &  2.50   &  2.7   &  6.65 \\
9.183  & 48190.2  &   1.44   &  2.80   &  2.7   &  6.75 \\
9.366  & 47983.4  &   2.90   &  3.83   & 10.0   & 12.03 \\
9.551  & 43828.8  &   4.51   &  5.09   & 10.0   & 12.51 \\
9.735  & 46788.7  &   3.72   &  4.45   & 10.0   & 12.25 \\
9.921  & 50939.4  &   0.94   &  2.52   &  2.7   &  6.66 \\
10.107 & 52751.5  &   1.02   &  2.54   &  2.7   &  6.68 \\
10.294 & 52128.7  &   0.82   &  2.51   &  2.7   &  6.65 \\
10.481 & 54768.2  &   0.64   &  2.47   &  2.7   &  6.63 \\
10.668 & 55583.9  &   0.95   &  2.61   &  2.7   &  6.67 \\
10.857 & 56168.0  &   1.00   &  2.56   &  2.7   &  6.67 \\
11.046 & 55322.7  &   2.06   &  3.22   &  2.7   &  6.91 \\
11.235 & 53247.8  &   3.50   &  4.19   &  2.7   &  7.47 \\
11.425 & 52871.8  &   1.28   &  2.68   &  2.7   &  6.72 \\
11.616 & 52857.9  &   1.35   &  2.70   &  2.7   &  6.73 \\
11.807 & 50134.5  &   3.80   &  4.47   &  2.7   &  7.62 \\
11.999 & 49050.2  &   1.41   &  2.74   &  2.7   &  6.75 \\
12.191 & 44697.1  &   2.19   &  3.30   &  2.7   &  6.95 \\
12.384 & 40722.9  &   2.58   &  3.47   &  2.7   &  7.09 \\
12.577 & 38720.9  &   7.14   &  7.75   &  2.7   &  9.72 \\
12.771 & 33237.0  &   5.53   &  6.46   &  2.7   &  8.61 \\
12.966 & 33431.4  &   1.61   &  3.72   &  2.7   &  6.79 \\
13.161 & 30643.0  &   3.92   &  5.16   &  2.7   &  7.67 \\
13.357 & 31091.5  &   6.04   &  6.91   &  2.7   &  8.95 \\
\enddata
\tablenotetext{1} {Measurement uncertainty, including IRC +10216 and the standard $\mu$ UMa.}
\tablenotetext{2} {Total local uncertainty, including 2.31-3.35\% uncertainty from the $\mu$ UMa spectrum of \citet{1996AJ....112.2274C}.}
\tablenotetext{3} {The estimated global telluric calibration uncertainty.}
\tablenotetext{4} {Includes 5\% systematic uncertainty from the SLCF.}

\end{deluxetable}

\end{document}